\documentclass[twocolumn]{aastex62}
\usepackage{amsmath,amstext}
\usepackage[figure,figure*]{hypcap}
\usepackage{newtxmath} 
\usepackage{version}
\usepackage{tikz}
\usetikzlibrary{shapes.geometric, arrows}
\tikzstyle{startstop} = [rectangle, rounded corners, minimum width=2cm, minimum height=1cm,text centered, draw=black, fill=red!30]
\tikzstyle{io} = [trapezium, trapezium left angle=70, trapezium right angle=110, minimum width=1cm, minimum height=1cm, text centered, draw=black, fill=blue!30]
\tikzstyle{process} = [rectangle, minimum width=1cm, minimum height=1cm, text centered, draw=black, fill=orange!30]
\tikzstyle{model} = [rectangle, minimum width=1cm, minimum height=1cm, text centered, draw=black, fill=pink!30]
\tikzstyle{decision} = [diamond, minimum width=1cm, minimum height=0.5cm, text centered, draw=black, fill=green!30]
\tikzstyle{arrow} = [thick,->,>=stealth]

\newcommand{\nicer}{\textit{NICER}}

\newcommand{\Chandra}{{\em Chandra}}
\newcommand{\chandra}{\textit{Chandra}}
\newcommand{\gaia}{\textit{Gaia}}
\newcommand{\chandralong}{\textit{Chandra X-ray Observatory}}
\newcommand{\xmmlong}{\textit{XMM-Newton}}
\newcommand{\xmm}{\textit{XMM}}

\newcommand{\nh}{\mbox{$N_{\rm H}$}}

\newcommand{\kteff}{\mbox{$kT_{\rm eff}$}}
\newcommand{\rinfty}{\mbox{$R_{\infty}$}}
\newcommand{\rns}{\mbox{$R_{\rm NS}$}}
\newcommand{\mns}{\mbox{$M_{\rm NS}$}}

\newcommand{\chisq}{\mbox{$\chi^2$}}
\newcommand{\chisqnu}{\mbox{$\chi^2_\nu$}}

\newcommand{\halpha}{\mbox{H$\alpha$}}

\newcommand{\mr}{\mbox{\mns--\rns}}

\newcommand{\simlt}{\mathrel{\hbox{\rlap{\hbox{\lower4pt\hbox{$\sim$}}}\hbox{$<$}}}}
\newcommand{\simgt}{\mathrel{\hbox{\rlap{\hbox{\lower4pt\hbox{$\sim$}}}\hbox{$>$}}}}

\newcommand{\ee}[1]{\mbox{$10^{#1}$}}
\newcommand{\tee}[1]{\mbox{$\times 10^{#1}$}}
\newcommand{\ud}[2]{\mbox{$^{+ #1}_{- #2}$}}
\newcommand{\ppm}{\mbox{$\pm$}}

\newcommand{\unit}[1]{\mbox{$\rm\,#1$}}

\def\sec{\mbox{$\,{\rm sec}$}}
\newcommand{\G}{\mbox{$\,G$}}

\newcommand{\msun}{\mbox{$\,{\rm M}_\odot$}}

\newcommand{\km}{\hbox{$\,{\rm km}$}}

\newcommand{\cm}{\mbox{$\,{\rm cm}$}}
\newcommand{\MeV}{\mbox{$\,{\rm MeV}$}}
\newcommand{\keV}{\mbox{$\,{\rm keV}$}}

\newcommand{\kpc}{\mbox{$\,{\rm kpc}$}}

\newcommand{\percmcube}{\mbox{$\,{\rm cm^{-3}}$}}
\newcommand{\peryear}{\mbox{$\,{\rm yr^{-1}}$}}

\newcommand{\cgsdensity}{\mbox{$\,{\rm g\percmcube}$}}

\def\OmCen{\mbox{$\omega$\,Cen}}

\definecolor{blue-munsell}{rgb}{0.0, 0.5, 0.69}
\definecolor{blush}{rgb}{0.87, 0.36, 0.51}
\definecolor{beaver}{rgb}{0.62, 0.51, 0.44}
\definecolor{asparagus}{rgb}{0.53, 0.66, 0.42}
\definecolor{antiquefuchsia}{rgb}{0.57, 0.36, 0.51}

\shorttitle{Equation of state constraints from NS thermal emission}
\shortauthors{Baillot d'Etivaux et al.}

\begin{document}

\title{New constraints on the nuclear equation of state from the thermal emission of neutron stars in quiescent low-mass X-ray binaries}

\author{Nicolas Baillot d'Etivaux}
\affil{Univ Lyon, Universit\'{e} Claude Bernard Lyon 1, CNRS/IN2P3, Institut de Physique Nucl\'eaire de Lyon, F-69622 Villeurbanne, France}

\author[0000-0002-6449-106X]{Sebastien Guillot} 
\correspondingauthor{Sebastien Guillot}
\thanks{sebastien.guillot@irap.omp.eu}
\affil{CNRS, IRAP, 9 avenue du Colonel Roche, BP 44346, F-31028 Toulouse Cedex 4, France}

\author[0000-0001-8743-3092]{J\'er\^ome Margueron}
\affil{Univ Lyon, Universit\'{e} Claude Bernard Lyon 1, CNRS/IN2P3, Institut de Physique Nucl\'eaire de Lyon, F-69622 Villeurbanne, France}
\affil{Institute for Nuclear Theory, University of Washington, Seattle, Washington 98195, USA}

\author{Natalie Webb}
\affil{CNRS, IRAP, 9 avenue du Colonel Roche, BP 44346, F-31028 Toulouse Cedex 4, France}

\author[0000-0001-6003-8877]{M\'{a}rcio Catelan}
\affil{Instituto de Astrof\'{i}sica, Facultad de F\'{i}sica, Pontificia Universidad Cat\'{o}lica de Chile, Av. Vicu\~{n}a Mackenna 4860,\\ 7820436 Macul, Santiago, Chile}
\affil{Instituto Milenio de Astrof\'{i}sica, Santiago, Chile}   

\author[0000-0003-4059-6796]{Andreas Reisenegger}
\affil{Instituto de Astrof\'{i}sica, Facultad de F\'{i}sica, Pontificia Universidad Cat\'{o}lica de Chile, Av. Vicu\~{n}a Mackenna 4860,\\ 7820436
   Macul, Santiago, Chile}

\begin{abstract}
This paper presents a new analysis of the thermal emission from the neutron star (NS) surface to constrain the dense matter equation of state. We employ an empirical parameterization of the equation of state with a Markov-Chain Monte Carlo approach to consistently fit the spectra of quiescent low-mass X-ray binaries in globular clusters with well-measured distances. Despite previous analyses predicting low NS radii, we show that it is possible to reconcile the astrophysical data with nuclear physics knowledge, with or without including a prior on the slope of the symmetry energy $L_{\rm sym}$. With this empirical parameterization of the equation of state, we obtain radii of the order of about 12~km without worsening the fit statistic. More importantly, we obtain the following values for the slope of the symmetry energy, its curvature $K_{\rm sym}$, and the isoscalar skewness parameter $Q_{\rm sat}$: $L_{\rm sym}=37.2^{+9.2}_{-8.9}$~MeV, $K_{\rm sym}=-85^{+82}_{-70}$~MeV, and $Q_{\rm sat}=318^{+673}_{-366}$~MeV.  These are the first measurements of the empirical parameters $K_{\rm sym}$ and $Q_{\rm sat}$. Their values are only weakly impacted by our assumptions, such as the distances or the number of free empirical parameters, provided the latter are taken within a reasonable range. We also study the weak sensitivity of our results to the set of sources analyzed, and we identify a group of sources that dominates the constraints. The resulting masses and radii obtained from this empirical parameterization are also compared to other measurements from electromagnetic observations of NSs and gravitational wave signals from the NS-NS merger GW~170817.
\end{abstract}


\section{Introduction}
 \label{sec:intro}

Determining the equation of state of dense matter (EoS) -- the relation between the pressure $P$ and energy-density $\rho$ beyond the nuclear saturation energy-density $\rho_{\rm sat} \sim 2.4\tee{14}\cgsdensity$ -- is an important goal of fundamental physics and astrophysics, with far-reaching implications. Observations of neutron stars (NSs) offer extraordinary tools to investigate dense matter properties, which are complementary to experimental studies~\citep[e.g.,][]{lattimer07,kramer08,lattimer10,lattimer2013,hebeler13}. For instance, macroscopic properties of NSs, such as masses, radii, moments of inertia or tidal deformabilities, provide constraints on dense matter at energy-densities beyond $\rho_{\rm sat}$ \citep[e.g.,][]{lattimer90,lattimer05,Flanagan2008,lattimer10,abbott18}.

A variety of methods exist to constrain the EoS from NSs. Besides electromagnetic observations described below, the recent observation of the gravitational wave signal from a NS-NS merger and its electromagnetic counterpart has been analyzed to better constrain the stiffness of matter inside NSs. Specifically, the signal GW~170817, detected by the LIGO and Virgo gravitational wave (GW) detectors on 2017 August 17th, resulted in constraints on the tidal deformability of the NSs from the quadrupole moment in the space-time surrounding the NS merger \citep{abbott17}. Following the discovery of GW~170817, several articles proposed constraints on the EoS and the radius of these NSs using information from the GW signal and the simultaneous GRB~170817, including its afterglow AT~2017gfo (e.g., \citealt{bauswein17,radice18,annala18,raithel18,Tews2018,de18}, with the most recent one from \citealt{abbott18}). The conclusions of these papers are consistent, although the real quantitative information extracted from this first ever detection may not yet compete with nuclear physics knowledge~\citep{Tews2018}. The future of this detection method is however  promising and will certainly constrain present EoS models.

All other methods to constrain the EoS make use of electromagnetic observations of NSs. More generally, they rely on mass \mns\ and radius \rns\ measurements (or other related properties). For example, the modelling of the pulse profile of millisecond pulsars (MSP) can provide measurements of \mns\ and \rns\ \citep[e.g.,][]{bogdanov07,bogdanov13}. The currently operating \textit{Neutron Star Interior Composition ExploreR} (\nicer) is routinely observing MSPs with this aim \citep{gendreau16,gendreau17}. Measuring the moment of inertia of pulsars using radio timing observations of pulsars in binary systems via spin-orbit coupling effects is also being envisaged to constrain the EoS \citep[e.g.,][]{lattimer05,kramer08}. Finally, the thermal emission from NSs  provides a promising technique to obtain \mns\ and \rns{} \citep[see ][for recent reviews]{miller13,heinke13,ozel16b}.  While this could, in principle, be achieved with all cooling NSs, some of them may be affected by systematic uncertainties that may alter the measurements.  For example, the spectral modeling of X-ray dim isolated NSs may be complicated by uncertainties about their atmospheres, their magnetic field $B\sim\ee{11-12}\G$, and the presence of X-ray pulsations indicating a non-uniform surface emission \citep[e.g.,][]{pons02}; which may require phase-resolved spectroscopy \citep{hambaryan17}. Similarly, central compact objects are likely affected by the same effects, although not all CCOs show pulsations \citep[e.g.,][]{klochkov15}.

The cooling tails of Type-I bursts from NSs in X-ray binaries have also been used for EoS constraints \citep[e.g.,][]{suleimanov11b,nattila16}. Furthermore, when these bursts reach the Eddington flux, the peak flux provides an additional observable with which to break the degeneracy between \mns\ and \rns{} \citep[e.g.][]{ozel10,guver13}. However, difficulties may arise from the spectral modeling with a Planck function and the use of a color correction from theoretical atmosphere models \citep[see ][for discussions]{guver12a,guver12b,kajava14,nattila16,ozel16a}. To remedy these issues, recent work fitted such atmosphere models to each spectrum during the cooling tail of the NS 4U~1702--429 to obtain \mns\ and \rns\ measurements \citep{nattila17}, instead of relying on color corrections.

All methods using thermally emitting NSs require precise knowledge of the source distances. For this reason, quiescent low-mass X-ray binaries (qLMXBs) located inside globular clusters (GCs) have provided reliable constraints on the EoS. The distances to GCs can be measured independently with uncertainties of $\sim$5--10\% \citep{harris96,harris10}; compared to the $\sim$ 30--50\% uncertainties of LMXBs in the field of the Galaxy. Furthermore, qLMXBs present other advantages that we describe in Section~\ref{sec:thermal}. While initially these sources where analyzed individually to constrain the EoS \cite[e.g.,][]{heinke06a,webb07,guillot11a}, it has become clear in recent years that statistical analyses combining multiple qLMXBs would provide more useful constraints on dense matter \citep{guillot13,guillot14,guillot16b,lattimer14,ozel16a,bogdanov16,steiner18}.

This article presents one such analysis in which the spectra of a sample of qLMXBs are simultaneously analyzed to constrain the EoS. Because the red-shifted radius, measured from the modelling of the observed spectrum, depends on both the gravitational mass and the physical radius, a simultaneous analysis of several qLMXB sources can help break degeneracies between these two properties of NSs, assuming these objects are governed by the same \mr\ relation, i.e., the same EoS. This can in turn be used to infer the properties of the dense NSs matter. For practical reasons, such method requires parameterizing the EoS, i.e., representing it as a function of some parameters either in \mr\ space, or in $P$--$\rho$ space.  Previous work used analytical parameterizations, such as a toy-model constant-\rns\ \citep{guillot13,guillot14,guillot16b} or piecewise polytrope representations\footnote{A sequence of connected power laws, $P=k_i\rho^{\gamma_i}$, where $i$ typically runs up to 3 or 5 \citep[e.g.,][]{read09,raithel16}.}~\citep{lattimer14,ozel16a,bogdanov16,steiner18}.

In this work, we employ a representation of the EoS based on nuclear physics empirical parameters. The model is presented in \cite{margueron18a,margueron18b} and offers the possibility to easily incorporate nuclear physics knowledge.  In Section~\ref{sec:thermal}, we summarize the characteristics of qLMXBs and present the reasons that make them ideal sources for EoS constraints. We also describe the data reduction and spectral extraction of our qLMXBs sample, as well as the surface emission model of these NSs.  Section~\ref{sec:eos} summarizes various aspects of the EoS meta-model of \cite{margueron18a,margueron18b} that we used to fit our spectral data of qLMXBs. Section~\ref{sec:mcmc} presents the Markov-Chain Monte Carlo (MCMC) approach used to find the best fit EoS model to the qLMXBs spectra  and Section~\ref{sec:results} presents the results, and compares them with previous constraints on the EoS. Finally, the conclusions in Section~\ref{sec:conclusions} summarize this work.

\section{Thermal emission from quiescent low-mass X-ray binaries}
\label{sec:thermal}

In this section, we detail our present understanding of qLMXB thermal emission in GCs as well as host GC distance measurements. We also give details on our X-ray spectral data analysis and spectral model.

\subsection{Low-mass X-ray binaries in quiescence}
\label{sec:qlmxb}

The surface emission from NSs in qLMXBs is now routinely used to obtain measurements of \mns\ and \rns.  While during outbursts the accreted matter dominates the X-ray emission, the thermal emission from the surface of the NS becomes visible in quiescence. The source of this thermal emission is internal, and originates from the heat deposited by nuclear reactions in the crust during accretion episodes \citep[e.g.,][]{haensel08}. As this emission, reprocessed by the NS outer layers, is observed in the X-rays and modeled with realistic atmosphere models \citep{zavlin96,heinke06a,ho09,haakonsen12}, one can measure the red-shifted temperature and the size of the emission area. In this way, the X-ray spectra of qLMXBs provide a measurement of \rinfty, defined as:
\begin{equation}
    \rinfty = \rns \left(1+z\right) = \rns \left(1-\frac{2 G \mns}{\rns c^{2}}\right)^{-1/2}.
\end{equation}
This requires knowing the distance to the source, and  qLMXBs located in GCs have provided \rns\ measurements since their distances can be independently and rather precisely measured (see Section~\ref{sec:distances}).

The qLMXBs inside GCs also present the other advantage of exhibiting a remarkable flux stability at all timescales \citep{heinke06a,guillot11a,servillat12,heinke14}. While LMXBs in the field of the Galaxy often exhibit flux variability, attributed to changes in the non-thermal and/or thermal components \citep[e.g.,][]{rutledge02a,campana04a}, which complicate the spectral modeling, the spectra of known qLMXBs located in GCs are purely thermal, without signs of non-thermal emission \citep[e.g.,][]{guillot13}. Overall, this reinforces the scenario in which we are observing the uncontaminated thermal cooling of NSs.

\begin{deluxetable*}{lccccccrr}
\tablecaption{Observational information on the 7 qLMXB sources considered in our analysis. \label{tab:sources}}
\tablehead{
\colhead{Globular} & \colhead{R.A.\tablenotemark{a}} & \colhead{Decl.\tablenotemark{a}} & \colhead{XMM Exp.} & \colhead{Chandra Exp.} & \colhead{S/N} & \colhead{Group\tablenotemark{b}} & \colhead{Distances} & \colhead{Distances [8]} \\
\colhead{Cluster host}& \colhead{(J2000)} & \colhead{(J2000)} & \colhead{time (ks)} & \colhead{time (ks)} & & &  \colhead{\emph{Dist \#1} (kpc)} & \colhead{\emph{Dist \#2} (kpc)}
} 
\startdata
47Tuc (X-7) & 00:24:03.53 & --72:04:52.2 & 0 & 181  & 122 & A,A' & $4.53 \pm 0.08$ [1] & $4.50 \pm 0.06$ \\
M28       & 18:24:32.84 & --24:52:08.4 & 0 & 327  & 113 & A,A' & $5.5 \pm 0.3$   [2,3] & $5.50 \pm 0.13$ \\
NGC~6397 & 17:40:41.50 & --53:40:04.6 & 0 & 340  & 82  & A,A' & $2.51 \pm 0.07 $ [4]  & $2.30 \pm 0.05$\\
\OmCen{}  & 13:26:19.78 & --47:29:10.9 & 36 & 291 & 49  & B,B' & $4.59 \pm 0.08$ [5] & $5.20 \pm 0.09$ \\
M13       & 16:41:43.75 & +36:27:57.7 & 29 & 55   & 36  & B,A' & $7.1 \pm 0.62$  [6] & $7.10\pm0.10$ \\
M30       & 21:40:22.16 & --23:10:45.9 & 0 & 49   & 32  & B,B' & $8.2 \pm0.62$ [6] & $8.10 \pm 0.12$ \\
NGC~6304 & 17:14:32.96 & --29:27:48.1 & 0 & 97   & 28  & B,B' & $6.22 \pm 0.26$ [7] & $5.90 \pm 0.14$ \\
\enddata
\tablenotetext{a}{Coordinates of the qLMXB in each of the GC.}
\tablenotetext{b}{The groups A and B denote the sources with a high S/N ($>60$) and lower S/N ($<60$), respectively. The groups A' and B' denote the sources for which we obtain a peaked and flat posterior distribution of the NS mass, respectively (see Section~\ref{sec:results} for more details).}
\tablecomments{All distance uncertainties are given at 1$\sigma$ confidence level. References: [1] \cite{bogdanov16}; [2] \cite{harris10} (with uncertainties estimated in [3] \citealt{servillat12}); [4] \cite{heinke14}; [5] \cite{watkins13}; [6] \cite{omalley17}; [7] \cite{recioblanco05};   [8] \cite{gaia18}, from which the distances were obtained from the individual $X$, $Y$, $Z$ coordinate values, as given in their Table~C.3, using $r_{\rm GC,\odot} = \sqrt{X^2+Y^2+Z^2}$.}
\end{deluxetable*}

Another advantage of NSs in qLMXBs over other sub-groups of NSs for the purpose of radius measurements is the relatively straightforward modeling of their emergent spectra.  While the atmospheric composition of isolated NSs may be uncertain \citep[e.g.,][]{burwitz03,ho09}, the atmosphere of NSs in LMXBs consists of a single-composition layer of a fully ionized light element. Since the accreted matter settles gravitationally within 10--100\sec\ \citep{alcock80,bildsten92}, the outermost layer of a transiently accreting NS is thought to be composed of the lightest accreted element, usually hydrogen (H).  Moreover, the magnetic fields of these old sources is thought to be weak, as supported by the fact that their presumed descendants, millisecond pulsars \citep{alpar82,bhattacharya91,tauris06}, have inferred dipole fields $B\sim10^{8}$--$10^{9}\G$, compared to $10^{11}$--$10^{12}$\G\ for the younger, ``classical'' pulsars, which have not undergone accretion. Such low $B$-fields do not affect the emergent spectrum, and it can therefore be assumed that the NS atmosphere is non-magnetic. For these reasons, H-atmosphere models, and in some cases Helium (He) atmosphere models (see below), have been used to fit the spectra of the NS in qLMXBs and extract measurements of \mns\ and \rns.

It is generally accepted that the atmosphere of a NS in a qLMXB is composed of pure H, since the atmospheric composition would be that of the lightest element present in the companion star. Unless the companion is completely devoid of H, the matter transferred onto the NS will contain some H, and therefore the material present in the outermost layer will be H. Diffusive burning of H into He may happen in the hot photosphere, but this is expected to happen on timescales of $10^{3}$--$10^{4}$ yrs \citep{chang04,chang10}, whereas the atmosphere (of thickness $\sim1\cm$ and mass $M_{\rm atm}\sim 10^{-20}\msun$, \citealt{bogdanov16}) can rapidly be replenished by H matter from the stellar companion, even at very low accretion rates of $\sim10^{-13}\unit{\msun\peryear}$.  More importantly, observational evidence demonstrated the presence of H in the  qLMXB systems in 47~Tux~X-5 \citep{edmonds02} and 47~Tuc~X-7~\citep{bogdanov16}, and in the GC \OmCen{} \citep{haggard04}. Searches for \halpha\ emission at the position of the qLMXB in NGC~6397 were unsuccessful, only placing upper limits on the  equivalent width of the spectral line and thus on the accretion rate \citep{heinke14}. It was therefore argued that this qLMXB was devoid of H. and the authors advocated a He atmosphere instead. This conclusion was supported by the low \rns\ found from the spectral analyses with a H atmosphere ($\sim 8\km$, in the earlier work of \citealt{guillot11a,guillot13}), while a He atmosphere resulted in a \rns\ value compatible with that of other NSs \citep{heinke14}. However, the stellar companion was only detected in the $R$-band, and limits on its photometric colors made it compatible with both the white-dwarf sequence and the main sequence of the host GC. However, as discussed below, the proper modeling of pile-up (an instrumental effect, \citealt{davis01}) in the \chandralong\ spectra of this qLMXB is sufficient to yield radii in the range\footnote{It was demonstrated that pile-up effects, even at the 1\%-level, can significantly shift the peak of the thermal spectrum to higher energies, and therefore result in underestimated radii \citep{bogdanov16}.} 10--11\km.

With all these considerations in mind, qLMXBs located inside GCs are ideal objects that provide a well-understood scenario to measure the radii of NSs. As mentioned above, obtaining constraints on the EoS from qLMXBs requires combining them into a statistical analysis. Here, we analyze the spectra of the qLMXB in the GCs M13 (NGC\,6205), M28 (NGC\,6266), M30 (NGC\,7099), NGC~6304, NGC~6397, \OmCen\ (NGC\,5139), and 47 Tuc~(NGC\,104)~X-7.  We excluded 47 Tuc~X-5 because of its eclipses, its flux variability, and variable line-of-sight absorption, which make the spectral modelling rather uncertain \citep{bogdanov16}.
Some information about these sources is detailed in Table~\ref{tab:sources}.

\subsection{On the distances of globular clusters}
\label{sec:distances}

In this paper, we work with a set of distances obtained from a heterogeneous set of methods, including dynamical  \citep{watkins13} and photometric (other references in Table~\ref{tab:sources} distance measurements. In most cases, these are recent measurements, or measurements discussed in previous qLMXB analyses, which we used for convenient comparison \citep[e.g.,][]{bogdanov16}. These distances used and their uncertainties are listed in Table~\ref{tab:sources} as \emph{Dist \#1}.

To evaluate the impact of the choice of distances, we also considered distances from a more uniform set of measurements. The determination of accurate astrometric distances to large samples of GCs have now become a tangible reality, thanks to the exquisite data provided by the European Space Agency's (ESA's) \gaia{} space mission \citep{gaia2016}. Within the framework of \gaia's Data Release 2 \citep[DR2;][]{gaiadr2}, trigonometric parallaxes have already become available for large numbers of stars belonging to dozens of GCs. Still, as discussed in detail by \citet{pancino-gaia} and more recently also emphasized by \citet{gaia-babusiaux}, systematic uncertainties still preclude the determination of reliable distances based on the available \gaia{} data for such crowded fields as Galactic GCs~-- even though, by the end of the mission, GC distances that are accurate to within the 1\% level can be expected \citep{pancino-gaia}. Confronting the \gaia-DR2 data with distances from the literature, as independently compiled in the \citet{harris96,harris10} catalog, a relatively small systematic offset, at the level of 0.029\,mas, was found \citep{gaia18}, in the sense that parallaxes derived by \gaia{} are smaller than those implied by the distances given in \citet{harris10}. In any case, at this stage, the \citet{gaia18,gaia-babusiaux} is using the latter distances, as opposed to those implied by the \gaia{} parallaxes, in its analyses of the Hertzsprung-Russell diagram and GC orbits.

Using the \citet{harris10} distances, the \cite{gaia18} rederived the $X$, $Y$, $Z$ coordinates of the GCs with respect to the Sun, given the improved positional information obtained by the \gaia{} mission. For our uniform set of distance measurements to the seven GCs studied here (\emph{Dist \#2}), we used the distances calculated from the $X$, $Y$, $Z$ coordinates in the \cite{gaia18}. We note that these distances are in most cases consistent with those of \emph{Dist \#1}, albeit with smaller uncertainties. The most significant difference between the two sets is for the GC \OmCen{}, although it has been noted that the dynamical measurement for this cluster \citep{watkins13} may suffer from systematics. Finally, we note that \cite{chen18} reported a distance to 47~Tuc of $4.45\pm0.01\pm0.12\kpc$ (statistical and systematic uncertainties) obtained from a careful treatment of the \gaia{}-DR2 parallaxes. This result is fully consistent with the values used in our sets \emph{Dist \#1} and \emph{Dist \#2}. Using these two sets allows us to study the impact of the distance choices on the analyses of X-ray spectra of thermally-emitting NSs.

\subsection{X-ray spectral data analysis and spectral model}
\label{sec:spec_model}

The processing of the \xmmlong\ and \chandra\ data sets is performed with the \emph{XMMSASv15.0} \citep{gabriel04} and \emph{CIAO v4.8} \citep{fruscione06}, respectively, following their respective standard procedures.  The spectra are created from flare-filtered event files, by extracting counts in circular regions. Background spectra are chosen from circular regions near the qLMXB, on the same CCD chip, and devoid of other sources. Finally, we grouped energy channels to ensure a minimum of 20 counts per bin.  A detailed description of the data preparation is available in \cite{guillot13}, and here we follow similar data reduction recipes.

The analysis of the qLMXB spectra is performed with \emph{PyXSPEC}, the Python interface to the fitting package \emph{XSPEC} \citep{arnaud96}. This allows us to employ an MCMC approach to sample the parameter space, as described in Section~\ref{sec:mcmc}.  The spectral model used is the NS H atmosphere model {\tt nsatmos} \citep{heinke06a}, modulated by absorption of soft X-rays by the interstellar medium.  For the Galactic absorption, we used the recent model {\tt tbabs} \citep{wilms00}.  We also add a power-law component to account for possible excess of counts above 2\keV\ that may originate from non-thermal emission. The exponent of this power law is fixed to 1.5, and we fit for the normalization.  As will be shown below, the contribution of this power-law component is consistent with being null for all qLMXBs.

A pile-up component is also added for all \chandra\ spectra, even those qLMXBs with low count rates inducing a pile-up fraction $\lesssim 1\%$.  As was pointed out by \cite{bogdanov16}, uncorrected pile-up, even at low pile-up fraction $\sim 1\%$, can significantly bias the radius measurement. Specifically, for NGC~6397, the low \rns\ obtained with H atmosphere models was a consequence of the unmodelled pile-up of photons in the X-ray spectra.

In summary, for each NS qLMXB in our sample, the spectral parameters of the model are:
\begin{itemize}
\item the parameter $\alpha$ in the {\tt pileup} model,
\item the column density of neutral hydrogen \nh, from the {\tt tbabs} model,
\item the NS surface temperature \kteff\ in the {\tt nsatmos} model,
\item the NS mass in the {\tt nsatmos} model,
\item the NS radius in the {\tt nsatmos} model,
\item the NS distance (set as a prior; see Table~\ref{tab:sources}) in the {\tt nsatmos} model,
\item the power-law normalization (model {\tt powerlaw} with fixed $\Gamma=1.5$).
\end{itemize}

In addition, multiplicative constants are used to account for absolute flux cross-calibration uncertainties between different detectors (\xmm-pn, \xmm-MOS, and \chandra). Therefore, for sources with spectra obtained with multiple detectors, multiplicative constants are added to the spectral model, as commonly done\footnote{In those case, the constant for \xmm-pn is fixed to unity, while the ones for the \xmm-MOS and \chandra\ spectra,  $C_1$ and $C_2$ respectively, are fitted parameters.}.  In this work, all NSs are assumed to be described by the same EoS. Therefore, their masses and radii will be tied together by the parameterized EoS described in the following section.

\section{The dense matter equation of state}
\label{sec:eos}

For the present analysis, the dense matter EoS is provided by a meta-modeling described in~\cite{margueron18a,margueron18b}, instead of the toy-model constant-\rns\ representation of the EoS \citep{guillot13,guillot14,guillot16b}, or instead of the polytropes \citep{steiner13,ozel16a,steiner18} used in previous works.  The meta-modeling employed here is able to accurately reproduce existing nucleonic EoSs and smoothly interpolate between them. It is based on a Taylor expansion in the baryon density $n=n_n+n_p$, where $n_n$ and $n_p$ are the neutron and proton densities, around the nuclear saturation density $n_\mathrm{sat}\approx 0.16$~fm$^{-3}$. Note that the nuclear saturation density is expressed as baryon number per unit volume and it coincides with the energy-density $\rho_\mathrm{sat}$ introduced previously. Such an approach is realistic up to 3--4 $n_{\rm sat}$, where one could expect the onset of new degrees of freedom (hyperons, quarks, pion condensation, etc). This meta-model may therefore break down for high-mass NSs (at around or above 2\msun). Fortunately, these high masses seem not to be favored in the present analysis and for the present sources. For completeness, we briefly describe our modeling for the crust and the core of the NSs in this section.

\subsection{Equation of state for cold catalyzed neutron stars}
\label{sec:eosmodel}

Our EoS spans from the outer crust of NSs down to their dense core. We consider the HP94 model for the outer crust, which represents it as a Coulomb lattice of spherical nuclei immersed in a gas of electrons \citep{haensel94}. In this model, the nuclear masses are the experimental ones when available, supplemented by a theoretical mass formula~\citep{moeller92} for the more exotic nuclei.  The inner crust starts when the energy density reaches $3.285\times 10^{11}$~g~cm$^{-3}$, and we consider the tabulated SLY EoS~\citep{douchin01} obtained from a Compressible Liquid Drop Model based on the Skyrme interaction SLy4~\citep{chabanat98}. A test of the sensitivity on the crust EoS can be performed by replacing the SLY EoS by another one, such as the FPS one. These two tabulated EoS can be downloaded from the following website\footnote{http://www.ioffe.ru/astro/NSG/NSEOS/}. 

For numerical reasons, the transition between the crust and the core is guided within and log$\rho$--log$P$ cubic spline matching the values and derivatives at both boundaries. The two boundaries are taken to be $n_{\rm sat}/10$ for the lower bound and $n_{\rm sat}$ for the upper one.  The sensitivity of this procedure to the choice of the boundaries is found to be small. Its impact on the total NS radius is less than 100~m, which is much smaller than current measurement uncertainties~\citep{margueron18b}.

In this work, we considered that the NS interior is made only of purely nucleonic matter, whose properties are obtained from the extrapolation of the known saturation properties of nuclear matter. These properties are encoded in the so-called empirical parameters of nuclear matter, which are defined as being the coefficients of the series expansion in terms of the density parameter $x=(n-n_{\rm sat})/(3n_{\rm sat})$ of the energy per particle in symmetric matter,
\begin{equation}
    e_{\rm sat} = E_{\rm sat}+\dfrac{1}{2}K_{\rm sat}x^2+\dfrac{1}{3!}Q_{\rm sat}x^3+\dfrac{1}{4!}Z_{\rm sat}x^4+...  \quad ,
\end{equation}
and of the symmetry energy per particle
\begin{equation}
    e_{\rm sym} = E_{\rm sym}+L_{\rm sym}x+\dfrac{1}{2}K_{\rm sym}x^2+\dfrac{1}{3!}Q_{\rm sym}x^3+\dfrac{1}{4!}Z_{\rm sym}x^4+... ,
\end{equation}
where the symmetry energy $e_{\rm sym}$ is defined as the isospin polarization energy
\begin{equation}
    e_{\rm sym} = \frac 1 2 \frac{\partial^2 e}{\partial \delta^2} \quad ,
\end{equation}
and where $\delta=(n_n-n_p)/(n_n+n_p)$ is the isospin asymmetry parameter and $e(n,\delta)$ is the nuclear energy per particle.

$E_{\rm sat}$ and $E_{\rm sym}$ are the saturation and symmetry energy at the saturation density $n_{\rm sat}$. $L_{\rm sym}$ is the slope of the symmetry energy, and since the saturation is an equilibrium point, there is no slope of the energy per particle in symmetric matter. $K_{\rm sat/sym}$ stands for the curvature, $Q_{\rm sat/sym}$ for the skewness, and $Z_{\rm sat/sym}$ for the kurtosis of the energy per particle in symmetric matter and of the symmetry energy, respectively. The values of these empirical parameters are determined from experimental measures,  with different accuracies. Reviews of their experimental determination can be found in~\cite{margueron18a} and in references therein.

\begin{deluxetable*}{lcccccccccccc}[t]
\centering
\tablecaption{Standard values and domain of variation of the empirical parameters considered in this analysis; taken from \cite{margueron18a}. See Section~\ref{sec:eosmodel} for the description of the parameters. \label{tab:empParam}}
\tablecolumns{13}
\tablewidth{0pt}
\tablehead{
\colhead{Emp. param.} & \colhead{$E_{\rm sat}$} & \colhead{$E_{\rm sym}$} & \colhead{$n_{\rm sat}$} & \colhead{$L_{\rm sym}$} & \colhead{$K_{\rm sat}$} & \colhead{$K_{\rm sym}$} & \colhead{$Q_{\rm sat}$} & \colhead{$Q_{\rm sym}$} & \colhead{$Z_{\rm sat}$} & \colhead{$Z_{\rm sym}$} & \colhead{$m^*$} & \colhead{$\Delta m^*$} \\
 & \colhead{(MeV)} & \colhead{(MeV)} & \colhead{(fm$^{-3}$)} & \colhead{(MeV)} & \colhead{(MeV)} & \colhead{(MeV)} & \colhead{(MeV)} & \colhead{(MeV)} & \colhead{(MeV)} & \colhead{(MeV)} & \colhead{($m_N$)} & \colhead{($m_N$)}
} 
\startdata
Standard  & -15.8 & 32.0 & 0.155 & 60 & 230 & -100 & 300 & 0 & -500 & -500 & 0.75 & 0.1 \\
Variation & -- & -- & -- & 20--90 &  --  & -400--200 & -1300--1900 &  --  &  --  &  --  &  --  & --  \\
\enddata
\end{deluxetable*}

We consider the meta-modeling ELFc proposed in~\cite{margueron18a}, which is based on the decomposition of the nuclear energy per particle in terms of a kinetic term $t$ and a potential term $v$, as
\begin{equation}
    e(n,\delta) = t(n,\delta)+v(n,\delta) \quad .
\end{equation} 
The kinetic energy is defined as that of the Fermi gas plus medium corrections to the bare mass (encoded in the parameters $\kappa_{\rm sat/sym}$),
\begin{equation}
    t(n,\delta)=\frac{t_{\rm sat}}{2}\left(\frac{n}{n_{\rm sat}}\right)^{2/3}\Big[\left(1+\kappa_{\rm sat}\frac{n}{n_{\rm sat}}\right)f_1(\delta)+\kappa_{\rm sym}\frac{n}{n_{\rm sat}}f_2(\delta)\Big],
\end{equation}

where $t_{\rm sat}=3\hbar^2/(10m)(3\pi^2/2)^{2/3}n_{\rm sat}^{2/3}$, $m$ is the nucleon mass, and the functions $f_{1/2}$ are defined as,
\begin{eqnarray}
    f_1(\delta) &=&(1+\delta)^{5/3}+(1-\delta)^{5/3},\\
    f_2(\delta) &=&\Big[(1+\delta)^{5/3}-(1-\delta)^{5/3}\Big]\delta.
\end{eqnarray}

The potential term is expressed as,
\begin{equation}
    v(n,\delta) = \sum^N_{\alpha=0} \left(v_\alpha^{\rm is} + \delta^2 v_\alpha^{\rm iv} \right)\frac{x^\alpha}{\alpha !} u(x),
\end{equation}
where the function $u(x)$ takes into account the corrections due to the truncation $N$ at low density, as
\begin{equation}
    u(x)=1-(-3x)^{N+1-\alpha}\exp(-b n/n_{\rm sat}).
\end{equation}
Fixing $b=10\ln 2\approx 6.93$, as in \cite{margueron18a}, implies that the function $u$ converges quickly to $1$ as the density increases from 0. It ensures that $v(n,\delta)\rightarrow 0$ for $n\rightarrow 0$ for any order $N$. The larger $N$, the smaller the correction $u(x)$. The parameters $v_\alpha^{\rm is/iv}$ entering into the series expansion of the potential term have a one-to-one relation with the empirical parameters.

The ability of this meta-modeling to reproduce existing EoS increases as the order $N$ increases. For $N=4$, the meta-modeling can very accurately (at the \% accuracy, in the worst case) reproduce binding energy, pressure, and sound velocity of a large number of existing EoS up to $4n_{\rm sat}$, as shown in \cite{margueron18a}.

In the present work, we use the flexibility of the meta-modeling to sample the parameter space of the empirical parameters using an MCMC approach. The range of variation for each of the empirical parameters considered in this analysis is given in Table~\ref{tab:empParam}. We fix the value of the lowest-order empirical parameters at saturation density to be: $E_{\rm sat}=-15.8$~MeV, $E_{\rm sym}=32$~MeV, $n_{\rm sat}=0.155$~fm$^{-3}$ and $K_{\rm sat}=230$~MeV. The parameters $\kappa_{\rm sat/sym}$ are adjusted so that the Landau mass in symmetric matter is $m^*/m=0.75$ and the splitting between the neutron and proton Landau masses $(m^*_n-m^*_p)/m$ in neutron matter is 0.1 (see Table~\ref{tab:empParam}). The \mr\ relation is known to be mostly influenced by the empirical parameters $L_{\rm sym}$, $K_{\rm sym}$, and $Q_{\rm sat}$, since the EoS in the density range going from $n_{\rm sat}$ to approximately $3n_{\rm sat}$ depends most strongly on them \citep{margueron18b}. $L_{\rm sym}$ and $K_{\rm sym}$ (respectively, $Q_{\rm sat}$) control the density dependence of the symmetry energy (respectively, the energy per particle in symmetric matter) above saturation density. The higher-order empirical parameters are poorly known, but they impact the EoS at higher densities. They could in principle be deduced from \mns\ and \rns\ measurements for high-mass NSs. 

\textsl{A priori}, we do not know which region of NS masses will be reached by our analysis. Anticipating our results, however, we find that the NS masses do not exceed 1.5--1.6~\msun, which implies that the central densities of these NSs are not very large, and the meta-model can reasonably be applied.

\subsection{The effect of the empirical parameters on the \mr\ relation}
\label{sec:nuclmodels}

We illustrate here the impact of the empirical parameters $L_{\rm sym}$, $K_{\rm sym}$ and $Q_{\rm sat}$ on the \mr\ relation. Since the rotation of the sources studied here is unknown, we consider non-rotating NS models, whose \mr\  relation for a given EoS is obtained by solving the well-known Tolman-Oppenheimer-Volkoff (TOV) equations \citep{tolman39,oppenheimer39}.  Only if the frequency is larger than 300 Hz (period\,$<3$~ms) the rotational effects could bias the \rns\ measurements \citep{morsink07}. For a NS with spin frequency of 600 Hz, its non-rotating radius would be underestimated by 2--5\%, depending on the NS size \citep{baubock13}. 

The EoS selection criteria include those which satisfy the requirements of causality and positiveness of the symmetry energy, as well as being compatible with a maximum mass above $1.9\msun$.  This mass limit corresponds approximately to the $2\sigma$ lower limits of the measurements for PSR~J1614--2230, $1.908\pm0.016\msun$ \citep{demorest10,fonseca16,arzoumanian18}, and PSR~J0348+0432 \citep{antoniadis13}, $2.01\pm0.04\msun$.

\begin{figure}[t]
\centering
\includegraphics[width=\columnwidth]{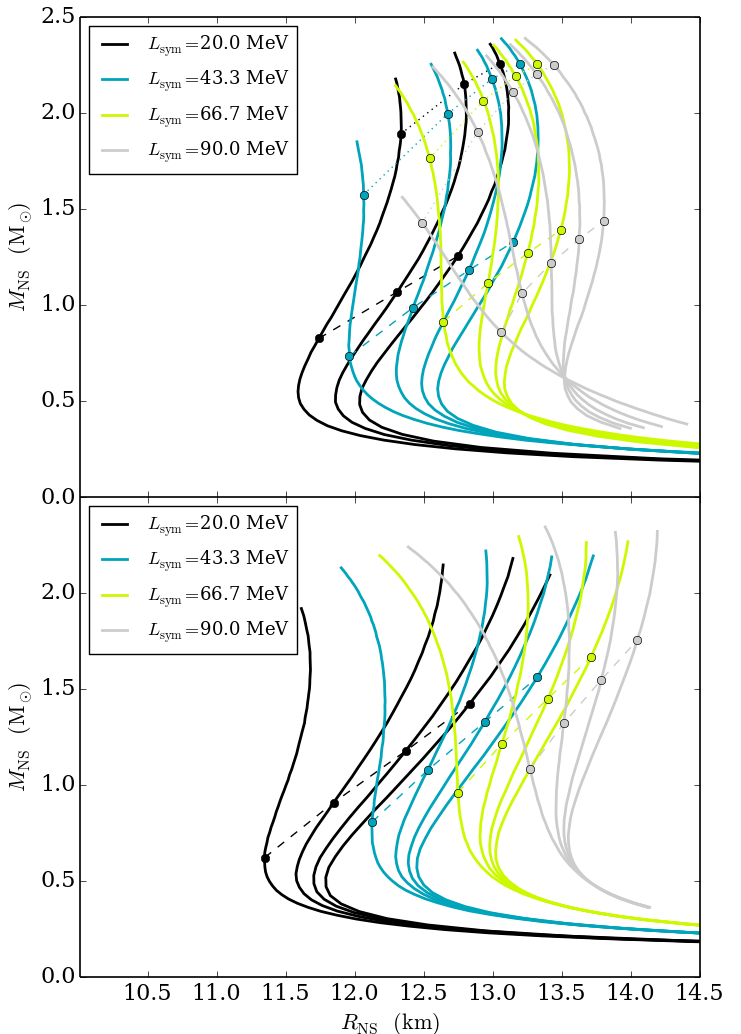}
\label{fig:MR1}
\vspace{-0.4cm}
\caption{This figure shows the effect on the \mr\ relations as the EoS parameters $L_{\rm sym}$, $K_{\rm sym}$ and $Q_{\rm sat}$ are varied. In the top panel, the value of $Q_{\rm sat}$ is fixed to 300 \MeV, while $L_{\rm sym}$ takes the values specified in the legend, and $K_{\rm sym}$ is varied from $-400\MeV$ to $200\MeV$ in steps of $120\MeV$ (increasing $K_{\rm sym}$ from left to right).  In some cases, the EoS for the lowest values for $K_{\rm sym}$ are not plotted if it does not match the selection criteria (see Section~\ref{sec:eos} for details). The points joined by the dashed and dotted line are models with central densities $2n_{sat}$ and $3n_{sat}$, respectively.  In the bottom panel, $L_{\rm sym}$ also takes the values specified in the legend, $K_{\rm sym}$ is fixed to -85\MeV, while $Q_{\rm sat}$ varies from $1900\MeV$ down to $-500\MeV$ in steps of -600\MeV. As in the top panel, the sets of the three parameters which do not satisfy the selection criteria are not plotted. Here, only the $2n_{sat}$ central densities points are shown, as some of the EoS displayed do not reach a central density of $3n_{sat}$.}
\end{figure}

Figure~\ref{fig:MR1} shows the effect of varying the empirical parameters $L_{\rm sym}$, $K_{\rm sym}$ and $Q_{\rm sat}$ on the \mr\ relation. Specifically, an increase of any of these three parameter shifts the high-mass part of the \mr\ relation to larger radii.  For clarity, only two parameters are varied in each of the top and bottom panels -- the third parameter being kept fixed ($Q_{\rm sat}=300\MeV$ in the top panel, and $K_{\rm sym}=-85\MeV$ in the bottom panel).  There are four groups of curves corresponding to the same value of $L_{\rm sym}$ and coinciding for very low mass NSs ($\mns<0.2\msun$). As \mns\ increases, the central density increases as well since we consider only the stable branch, and the different values for $K_{\rm sym}$ change the \mr\ curves associated with the different EoSs. Overall, varying $L_{\rm sym}$, $K_{\rm sym}$ and $Q_{\rm sat}$ over the whole range allowed by nuclear physics, together with the requirement of supporting a $1.9\msun$ NS, yields radii between 11.5 and 14.2~km at 1.4\msun.  The effect of varying the parameter $Q_{\rm sat}$ is most noticeable for \mns\ above 1.0--1.2\msun. Being of higher order in the density expansion, $Q_{\rm sat}$ influences the EoS at high density only, or equivalently at high \mns\ only. Depending on the value of $Q_{\rm sat}$, the EoS can be stiffer at high density, as reflected in the curves which go straight up, or softer at high density letting the \mr\ curve populate the low-\rns\ space at high \mns. There is however a limitation in the radii which can be explored based on the nucleonic EoS. As suggested in~\cite{margueron18b}, low-mass NSs with $\rns<11$~km (at $\sim 1.4\msun$) cannot be described by nucleonic EoS models that respect causality and that must support a 1.9\msun\ NS.

While there are various EoSs which pass through a point in the \mr\ diagram, their paths are different. The degeneracy between different EoSs thus requires the knowledge of a set of \mr\ points, as distant as possible from each other.  In conclusion of this analysis, the empirical parameters $L_{\rm sym}$, $K_{\rm sym}$, and $Q_{\rm sat}$ allow the exploration of a wide domain of \mns\ and \rns\ with various paths. Therefore, it may be possible to constrain the values of these parameters by confronting them to the observational data from the thermal X-ray emission of NSs.

\section{Confronting the equation of state with the data}
\label{sec:mcmc}

In this section, we detail the methodology of our analysis: we employ an MCMC approach with the stretch-move algorithm \citep{goodman10} to consistently analyze the seven qLMXB sources, and the nuclear matter EoS meta-modeling is included directly in the analysis. The result is that the astrophysical (NS properties) and nuclear physics (EoS) parameters are adjusted together, without over-constraining one or the other. We can solve the so-called inverse problem and obtain constraints on the EoS properties directly from the data analysis. This is the first time that the thermal emission from NSs is analyzed in this manner. 

\subsection{MCMC approach with the stretch-move algorithm}
\label{sec:gw}

For all the cases considered here, the priors on the parameters are chosen so as to minimize any \textsl{a priori} assumption on the parameter distributions. All astrophysical parameters (except the distances to the sources) are sampled with uniform distributions within the boundaries allowed by the spectral model (defined in \texttt{Xspec}). 

The distances $D$ are strongly coupled to the NS radii and effective surface temperatures. Letting this parameter explore a uniform prior would increase the uncertainties in the analysis enormously. For the two sets of distances presented in Table~\ref{tab:sources}, we limit the qLMXB distances to Gaussian priors given by the central values and $1\sigma$ uncertainties listed. To do so,  we add to the likelihood $\chi^2$ a penalty for each source $i$, proportional to the difference between the MCMC sampled distance $D_{\rm{mcmc,i}}$ and the actual measured data $D_{\rm{data,i}}$ (from Table~\ref{tab:sources}), taking into account their standard deviations $\sigma_i$. The distance penalty reads $\chi^2_D= \sum_{i=0}^{N}\chi^2_{D,i}$, where the $\chi^2_{D,i}$ for each source are given by:
\[  \chi^2_{D,i}=\dfrac{ \left( D_{\rm{mcmc,i}}-D_{\rm{data,i}}\right)^2  }{ \left(\sigma_i\right)^2} \, . \]

The MCMC approach permits efficient sampling of our parameter space with high dimensionality: 49 parameters in total, including 3 nuclear physics EoS parameters, plus 6 astrophysical parameters per qLMXB (those listed in Section~\ref{sec:spec_model}, except for the radii which are obtained given the sampled EoS parameters and NS masses, after solving the TOV equations), plus 4 multiplicative normalization constants (for the cross-calibration between the \xmm-pn, \xmm-MOS and \chandra, for the qLMXBs in M13 and \OmCen).

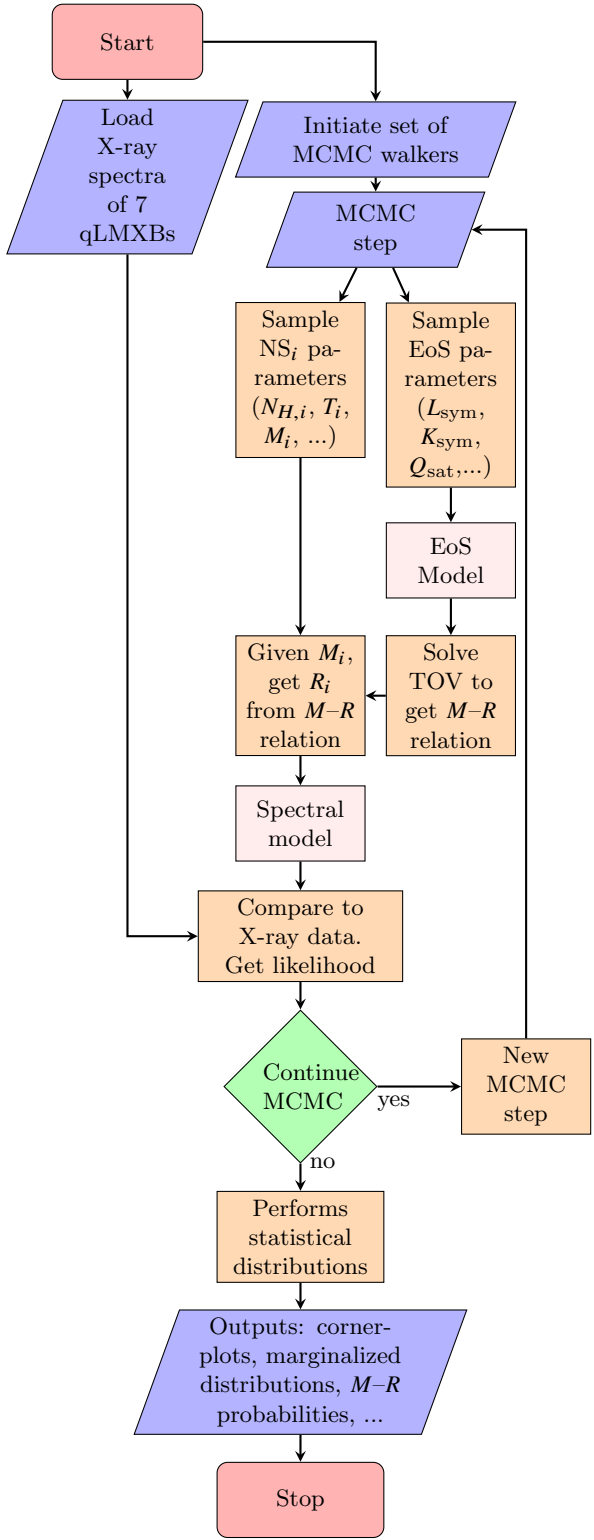
\begin{figure}
\centering
\begin{tikzpicture}[node distance=2cm]
\node (start) [startstop] {Start};
\node (in1) [io, below of=start, text width=1.5cm,yshift=0.2cm ] {Load X-ray spectra of 7 qLMXBs};
\node (in2) [io, right of=in1, yshift=0.5cm, xshift=1.3cm, text width=2.25cm] {Initiate set of MCMC walkers};
\node (in3) [io, below of=in2, yshift=0.8cm,text width=1.7cm] {MCMC step};
\node (proc1a) [process, below of=in3, xshift=-1cm, yshift=0.0cm, text width=1.5cm] {Sample NS$_{i}$ parameters ($N_{H,i}$, $T_{i}$, $M_{i}$, ...)};
\node (proc1b) [process, right of=proc1a, yshift=-0.2cm, text width=1.5cm] {Sample EoS parameters ($L_{\rm sym}$, $K_{\rm sym}$, $Q_{\rm sat}$,...)};
\node (proc1c) [model, below of=proc1b, text width=1.5cm, yshift=-0.2cm] {EoS Model};
\node (proc1d) [process, below of=proc1c, text width=1.5cm, yshift=0.2cm] {Solve TOV to get $M$--$R$ relation};
\node (proc1e) [process, left of=proc1d, text width=1.5cm] {Given $M_{i}$, get $R_{i}$ from $M$--$R$ relation};
\node (proc1f) [model, below of=proc1e, text width=1.5cm, yshift=0.3cm] {Spectral model};
\node (proc2)  [process, below of=proc1f, text width=2.5cm, yshift=0.5cm] {Compare to X-ray data. Get likelihood};
\node (dec1) [decision, below of=proc2, yshift=0.0cm, text width=1cm] {Continue MCMC};
\node (proc2a) [process, right of=dec1, xshift=1cm, text width=1.5cm] {New MCMC step};
\node (proc2b) [process, below of=dec1, text width=2cm, yshift=0.0cm] {Performs statistical distributions};
\node (out1) [io, below of=proc2b,text width=3cm, yshift=0.2cm] {Outputs: corner-plots, marginalized distributions, $M$--$R$ probabilities, ...};
\node (stop) [startstop, below of=out1,text width=2cm, yshift=0.3cm] {Stop};
\draw [arrow] (start) -- (in1);
\draw [arrow] (start) -| (in2);
\draw [arrow] (in2) -- (in3);
\draw [arrow] (in1) |- (proc2);
\draw [arrow] (in3) -- (proc1a);
\draw [arrow] (in3) -- (proc1b);
\draw [arrow] (proc1a) -- (proc1e);
\draw [arrow] (proc1b) -- (proc1c);
\draw [arrow] (proc1c) -- (proc1d);
\draw [arrow] (proc1d) -- (proc1e);
\draw [arrow] (proc1e) -- (proc1f);
\draw [arrow] (proc1f) -- (proc2);
\draw [arrow] (proc2) -- (dec1);
\draw [arrow] (dec1) -- node[anchor=north east] {yes} (proc2a);
\draw [arrow] (proc2a) |- (in3);
\draw [arrow] (dec1) -- node[anchor=south west] {no} (proc2b);
\draw [arrow] (proc2b) -- (out1);
\draw [arrow] (out1) -- (stop);
\end{tikzpicture}
\caption{Flowchart of the global fit to the data (X-ray spectra of 7 qLMXBs) with a set of walkers, and using MCMC and stretch-move algorithm (see text for more details). Note that the EoS model is implemented inside the observational analysis to provide consistent $MR$ relations.\label{fig:flowchart}}
\end{figure}

We use the python \texttt{emcee} package \citep{foremanmackey13} with the stretch-move algorithm \citep{goodman10}, which we applied as follows (see also the flow-chart in Figure~\ref{fig:flowchart}): 
\begin{itemize}
    \item Step 0: a large number of chains or "walkers" are initialized, each one corresponding to a random point in the multi-dimensional parameter space defined by the set of parameters described above. We use 426 walkers (a multiple of the number of CPU cores available for our study).
    \item Step 1: we solve the TOV equations for each walker, providing 426 \mr\ relations at each iteration.
    \item Step 2: for each walker, the sampled masses of the seven NSs are associated to seven calculated radii according to the \mr\ relation. Using those \mns\ and \rns\ and the other astrophysical parameters, we calculate the global $\chi^2$ between the emission models (NS atmosphere) and the data for the seven NSs. 
    \item Step 3: Given the calculated probability (its likelihood multiplied by the distance Gaussian priors mentioned above), the evolution of the walkers in the parameter space is decided according to the stretch-move algorithm. To determine the new position, each walker is randomly paired with another and will move along the line joining the two current points in the parameter space. The amount of "stretch" is determined by the scale-parameter (the only adjustable parameter in this algorithm), that have been chosen to $a=2.0$ as prescribed in \cite{goodman10}. The new position is accepted or rejected depending on its probability.  For more details about the stretch-move algorithm, see \cite{goodman10,foremanmackey13}.
    \item Step 4:  Steps 1 to 3 are repeated numerous times until the walkers have converged in the region of the parameter space resulting in the highest likelihood, or minimum $\chi^2$ value.
    \item Step 5: When the MCMC loop stops, the statistical posterior distributions are calculated and marginalized to create the outputs.
\end{itemize}

Before running the code on the full data set and with the most general meta-modeling in section~\ref{sec:results}, we first test it considering the constant-\rns\ toy-model. In addition to its simplicity, this test is interesting since it allows us to compare with results that have already been reported in the literature.

\subsection{Tests using a constant radius toy model}

We first consider the constant-\rns\ model \citep{guillot13}, which assumes that all NSs have the same radius, i.e., that the EoS is represented in \mr\ space by a vertical line in which \rns\ is independent of \mns\ (which remain as free parameters). This is a simple toy-model approximation motivated by the observations that most nucleonic EoSs (the ones consistent with 2\msun) have a rather weak dependence on \mns, between 1\msun\ and 2\msun. The purpose of this toy-model is mainly to test our code and MCMC approach.

\begin{figure}[t]
\centering
\includegraphics[width=\columnwidth]{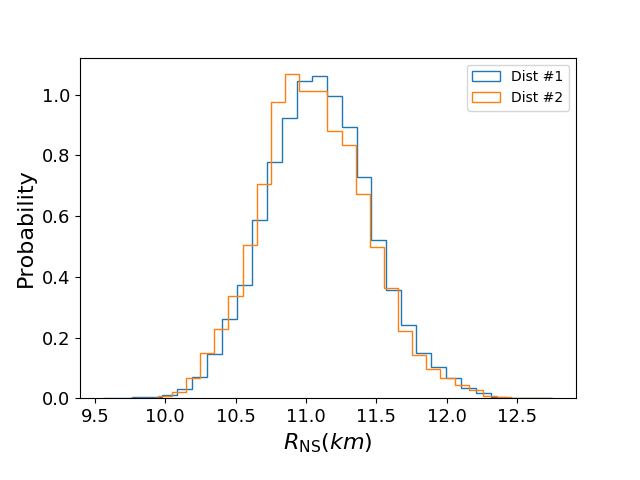}
\caption{(top) Marginalized posterior probability distributions of the radius obtained for the constant-\rns\ toy-model used in the MCMC tests runs (with the two sets of distances).}
\label{fig:radConst}
\end{figure}

After running the analysis described in Section~\ref{sec:gw}, we obtain the results shown in Figure~\ref{fig:radConst}, i.e., the \rns\ posterior distributions (the same for all seven NSs) considering the distance sets \emph{Dist \#1} and \emph{Dist \#2}, marginalized over the other parameters of the model. For both distance sets, the radius distributions are $\rns=11.09^{+0.38}_{-0.36}\km$ (\emph{Dist \#1}, for $\chisqnu=1.06$) and $\rns=11.04^{+0.39}_{-0.35}\km$ (\emph{Dist \#2}, for $\chisqnu=1.07$).  These values are consistent with the recent results of \cite{guillot16b}, but at odds with older results \citep[e.g.][]{guillot13, guillot14}. The differences are likely due to the inclusion of new sources (47Tuc~X-7) and new data, the use of recent distance measurements, the improvement of the analysis (e.g., the new absorption model \texttt{tbabs}), and the inclusion of the pile-up correction model for all sources (including those with a $\sim1\%$ pile-up fraction, see Section~\ref{sec:qlmxb}). Overall, we find a radius distribution that is easier to reconcile with the nuclear physics models of Section~\ref{sec:eos}, i.e., having non-negligible probabilities for a NS radius larger than about 11\km.

There is however also a large fraction of the posterior probability distribution which is located below 11\km, in conflict with nuclear physics expectations \cite[e.g.,][]{margueron18b,Tews2018} as well as our illustrative Figure~\ref{fig:MR1}. For instance, one could deduce from these figures that $\rns\lesssim 11\km$ requires $L_{\rm sym}\lesssim 20\MeV$, which contradicts nuclear physics expectations \citep{lattimer2013}.

In the following section, we address the question of the compatibility between the thermal emission modeling and the nuclear EoS by including the meta-model directly in the global spectral data analysis. In this way, we show that there is no inconsistency between the observational data and the nuclear EoS, and we extract an estimation for the nuclear EoS parameters.

\section{Results}
\label{sec:results}

In this section, the main results of our novel approach are presented and discussed.

\subsection{Framework}

We remind the reader that the main features of our work are that i) we fit simultaneously seven NS qLMXB sources, ii) we impose the same EoS to all these sources, and iii) we treat the EoS and the astrophysical model parameters equally.

Only a few nuclear EoS parameters are taken as free parameters. We recall that the nuclear meta-modeling is governed by a set of empirical parameters (see Section~\ref{sec:eos}). Some of these empirical parameters can be well constrained by nuclear experiments \citep[see the discussion in][]{margueron18a}, and they are kept fixed in the present analysis: $n_{\rm sat}$, $E_{\rm sat}$, $E_{\rm sym}$, and $K_{\rm sat}$ (values in Table~\ref{tab:empParam}). The more influential and less known parameters, $L_{\rm sym}$, $K_{\rm sym}$ and $Q_{\rm sat}$,  are fitted in our analysis. The values of $K_{\rm sym}$ and $Q_{\rm sat}$ are currently unknown, while there exist constraints on $L_{\rm sym}$ from nuclear experiments and nuclear theoretical predictions. These constraints indicate that $L_{\rm sym}$ has a value around 50~MeV with an uncertainty of about \ppm10~MeV \citep{lattimer2013}. We therefore incorporate this knowledge from nuclear physics by considering a Gaussian prior on $L_{\rm sym}$ centered at 50~MeV with a width of 10~MeV. Since $K_{\rm sym}$ and $Q_{\rm sat}$ are unknown, we consider a uniform distribution in the wide ranges listed in Table~\ref{tab:empParam}. The higher-order empirical parameters, $Q_{\rm sym}$ and $Z_{\rm sat/sym}$, are not known. However, since they influence only the high-density part of the EoS, they will not be tightly constrained by the present analysis. Therefore, they can be fixed to the values listed in Table~\ref{tab:empParam} \citep{margueron18b}.

\subsection{Main results}
\label{sec:mainresults}

The MCMC routine (described in Section \ref{sec:mcmc}) was run considering the seven qLMXB sources mentioned above. We have considered the chains that converged to the global minimum, excluding a few percent (1--5 \%) stuck at higher \chisq\ values (typically for reduced \chisq\ above 10). We tested the presence of these "stuck chains" with repeated iterations of the exact same MCMC run. In each case, the minimum best-fit \chisq\ is always found to be the same, and a small fraction of chains remain in the high-\chisq\ parts of the parameter space.  After 150,000 iterations, the reduced \chisq\ distribution is centered around $1.10\pm 0.02$, for 1126 degrees of freedom, and the best fit corresponds to $\chisq=1.08$, giving a null hypothesis probability of 3.1\%.

\begin{figure}
\centering
\includegraphics[width=\columnwidth]{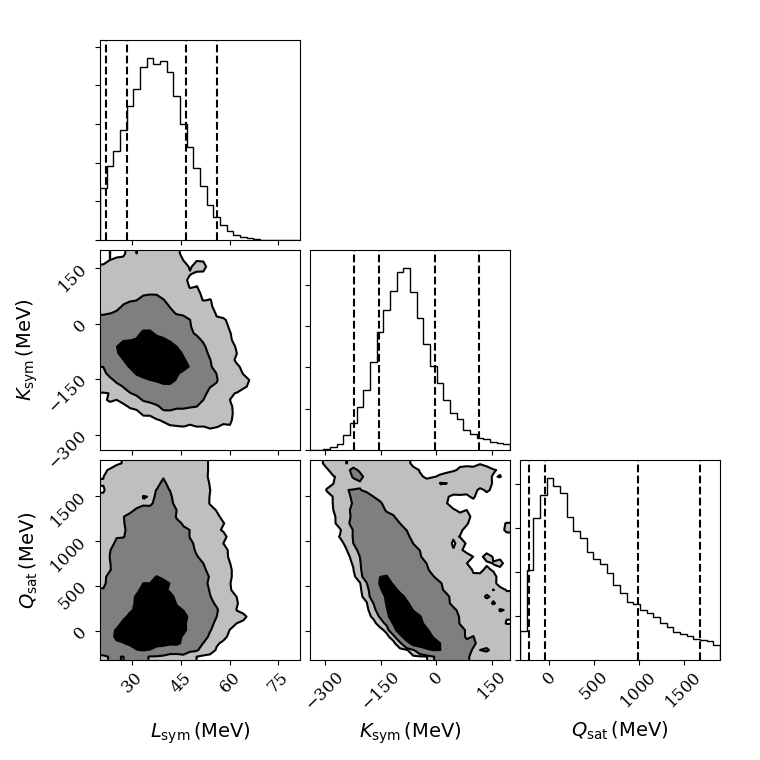}
\caption{Marginalized posterior probability distributions and correlations of the empirical parameters $L_{\rm sym}$, $K_{\rm sym}$ and $Q_{\rm sat}$. On the two-dimensional correlation plots, the contours indicate the 1, 2, and 3$\sigma$ confidence areas. On the one dimensional posterior distributions, the dashed vertical lines show the 68\% and 90\% quantiles around the median values. Here, all seven qLMXBs are included, the prior on $L_{\rm sym}=50\pm10\MeV$ is considered and the distances are determined from the set \emph{Dist \#2}. }
\label{fig:lkq}
\end{figure}

The marginalized posterior probabilities for the empirical EoS parameters $L_{\rm sym}$, $K_{\rm sym}$ and $Q_{\rm sat}$ are shown in Figure~\ref{fig:lkq}. We observed from the marginalized distributions that $L_{\rm sym}$ peaks at lower values than the one imposed by the prior ($50\pm10\MeV$), but remains consistent with it: $L_{\rm sym}=37.2^{+9.2}_{-8.9}\MeV$. This somewhat reflects the tension driving the fit towards low radii at low masses (see Figure~\ref{fig:MR1} and related discussion). The empirical parameter $K_{\rm sym}=-85^{+82}_{-70}\MeV$ is rather well constrained compared to the uniform prior, showing that this parameter is important for our data set.  Notice that it is also remarkably compatible with the one $-100\pm 100\MeV$ extracted from analysis of chiral effective field theory (EFT) calculations \citep{margueron18a}. Finally, the empirical parameter $Q_{\rm sat}=318^{+673}_{-366}\MeV$ is less constrained, but there is a preference for the lower values of the uniform prior distribution. The values of the empirical parameters for this run are reported in the first row of Table~\ref{tab:res1}. We point out that, despite the rather large uncertainties on the empirical parameters $K_{\rm sym}$ and $Q_{\rm sat}$, this is the first time that these parameters are extracted from data.

The correlations among empirical parameters are also visible in Figure~\ref{fig:lkq}. There is a weak anti-correlation between $L_{\rm sym}$ and $K_{\rm sym}$ and a stronger anti-correlation between $K_{\rm sym}$ and $Q_{\rm sat}$. These correlations reflect the causality and stability requirements, implying, for instance, that a large value for $K_{\rm sym}$ shall be compensated by a small value of $L_{\rm sym}$ or of $Q_{\rm sat}$ to limit the upper bound for the sound velocity, and vice-versa for the lower bound. The anti-correlation between $L_{\rm sym}$ and $K_{\rm sym}$ was already found in~\cite{margueron18b}, but the empirical parameters $L_{\rm sym}$/$K_{\rm sym}$ and $Q_{\rm sat}$ were found to be correlated for a stiff EoS (if the direct URCA process occurs for $\mns < 2\msun$) while for soft EoS (no direct URCA possible for $\mns < 2\msun$) no correlations were found. The anti-correlation between $K_{\rm sym}$ and $Q_{\rm sat}$ is therefore a new feature coming from the fit to the thermal x-ray emission.

\begin{figure}[t]
\centering
\includegraphics[width=\columnwidth]{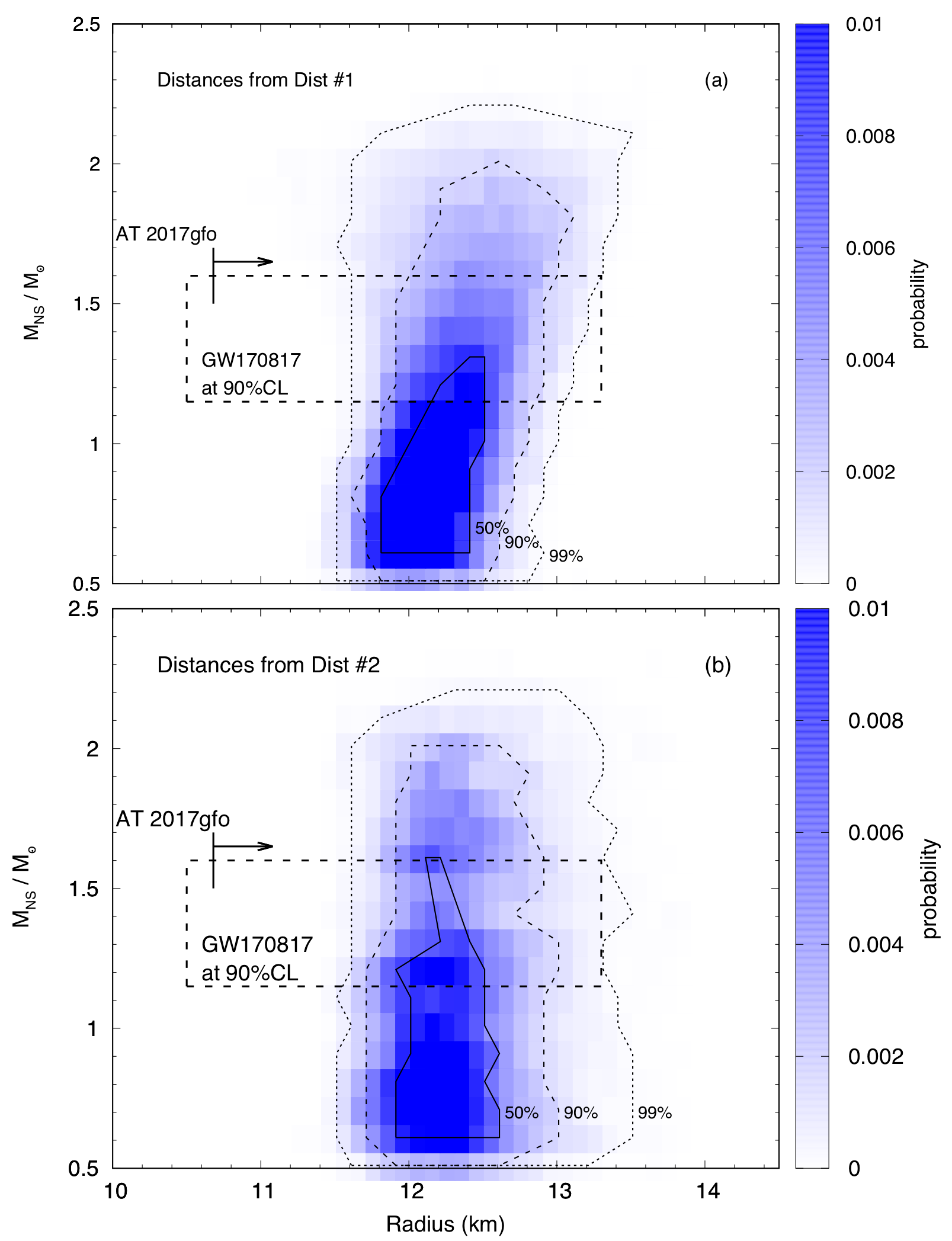}
\caption{\mr\ posterior probability distributions considering all the seven qLMXB sources, a prior on $L_{\rm sym}$ and on the distances from the set \emph{Dist \#1} (upper panel) and \emph{Dist \#2} (lower panel).  The 50\%, 90\% and 99\% confidence levels are represented, as well as the constraints from AT2017gfo \citep{bauswein17} and from GW170817 \citep{annala18,abbott18,Tews2018}.}
\label{fig:lkqMR}
\end{figure}

The \mr\ posterior probability distributions corresponding to the MCMC runs with the distances of set \emph{Dist \#1} (upper panel) and set \emph{Dist \#2} (lower panel) are displayed in Figure~\ref{fig:lkqMR}.  It is reassuring to notice that the \mr\ posterior probability distribution is almost insensitive to the set of distances considered, as was also observed for the constant-\rns\ test runs (Figure~\ref{fig:radConst}).  The global features of the probability distribution are the same for the two distance sets:  The radius that we obtain is between $\sim11.5$ and $13.0\km$ for a 1.4\msun\ NS, with a preference for low masses, although the 90\% credible intervals are compatible with 2\msun. 

\begin{figure*}[t]
\centering
\includegraphics[width=\textwidth]{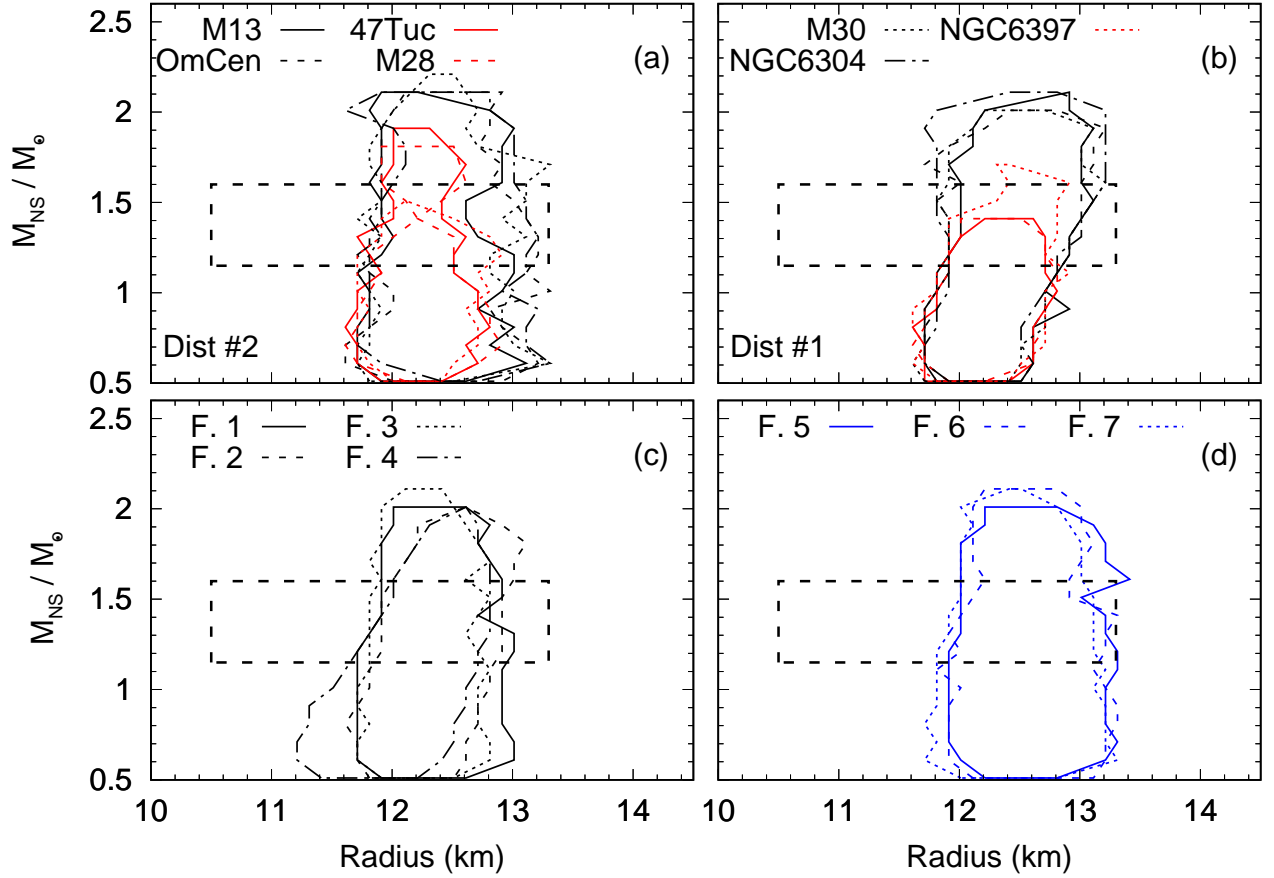}
\caption{({\it top}) 90\% credible contours of the posterior probability distributions in \mr\ space for the individual sources in the case of the Frameworks~\#1 (a) and \#2 (b). Note that for visibility, the legend indicating the individual sources is split between the panels (a) and (b).  ({\it bottom}) 90\% credible contours of the posterior probability distributions in \mr\ space for the Framework~\#1 to \#4 (panel c) and \#5 to \#7 (panel d). See Table~\ref{tab:res1} for the full results.  The constraints GW170817 \citep{abbott18} are also shown.}
\label{fig:indiv_fr1_7}
\end{figure*}

In Figure~\ref{fig:indiv_fr1_7} (top panels), we present the 90\% credible interval \mr\ posterior distributions of individual sources, obtained with the distance sets \emph{Dist \#2} (panel a) and \emph{Dist \#1} (panel b). Most sources have credible intervals that reach $\sim1.9\msun$ or higher. In a few cases, the 90\% credible intervals only reach masses around 1.4--1.5\msun, which appear to favor low-mass NSs. However, this is compatible with current distribution of MSP masses \citep{antoniadis16,ozel16b} which descend from LMXBs; we note that the lowest known NS mass is 1.174\ppm0.004\msun\ (for PSR~J0453+1559; \citealt{martinez15}).  Therefore, at the moment, there are no discrepancies with our current knowledge of NS formation mechanisms and their expected masses.

We note that the masses of the individual sources are not constrained as well as the radius (Figure~\ref{fig:indiv_fr1_7}). This is inherent to the method used in the present work, in which the measurable physical quantity is \rinfty. The constraints on the radius emerge from the combination of 1) the general shape in \mr-space of most EOS models in our nucleonic parameterization, and 2) the shape of the quantity \rinfty\ extracted from qLMXB spectra (see previous works, e.g., \citealt{guillot11a, bogdanov16, shaw18} for the \mr\ constraints from single qLMXBs, for which a significant portion of the \mr\ contours appear at constant \rns.). Therefore, unless the mass of the NS is measured independently, the degeneracy between \mns\ and \rns\ from observations of qLMXB can only be minimized by the implementation of an EOS parameterization as done in this work. Other events involving NSs enable measurements of \mns\ and \rns\ independently. For example, type I X-ray bursts with photospheric radius expansion bursts provide two observables (the Eddington flux and the cooling tail normalization, which both depend on \mns\ and \rns, see \citealt{ozel16b} for a review). The GW signals of two merging NSs, on the other hand, provides measurements of the merging masses and of the tidal deformability, which can be used to derive the radius \citep[e.g.,][]{abbott18,de18}.

The 50\%, 90\% and 99\% contours resulting from the present analysis can also be compared with the constraints from AT2017gfo \citep{bauswein17} and from GW170817 \citep{annala18,abbott18,Tews2018}.  There is a good agreement between these different constraints.  Nonetheless, the width of the distribution obtained from the present work appears narrower than those from analyses of GW~170817, indicative of more restrictive constraints on the NS radius, but this could also be due to the fact that we do not consider phase transitions in the meta-modeling of our work.

\begin{figure}[t]
\centering
\includegraphics[width=\columnwidth]{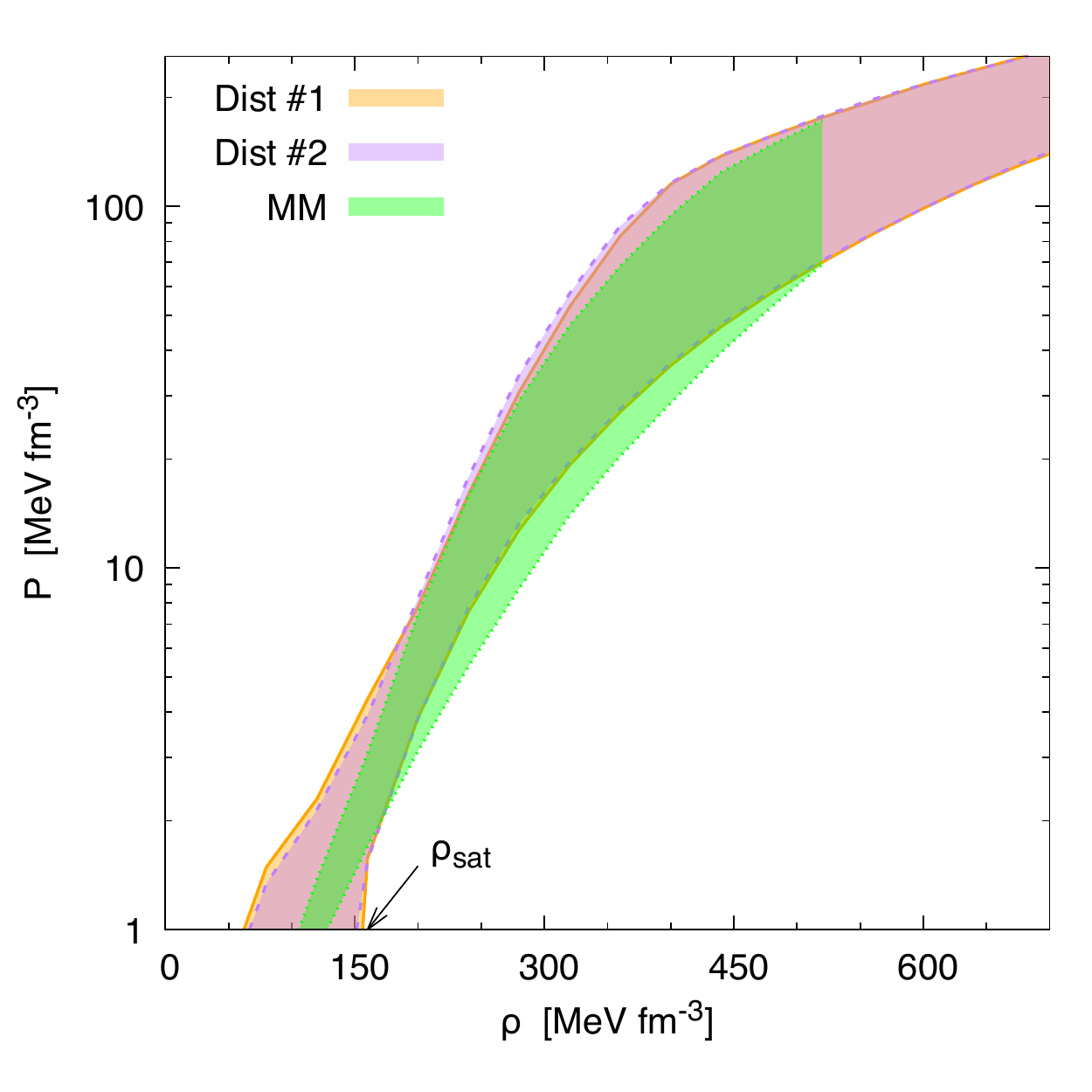}
\vspace{-0.8cm}
\caption{Boundary contours obtained for the pressure $P$ as a function of the energy density $\rho$, considering all the 7 qLMXB sources, the $L_{\rm sym}$ prior and the distances from \emph{Dist \#1} (orange band with solid contour) and \emph{Dist \#2} (purple band with dashed contour). The green band with dotted contour represents the prediction of the meta-model (MM) constrained by chiral EFT calculations in nuclear matter and the observed maximum mass of NS. There is a good overlap between the observed and the MM predictions for the EoS.}
\label{fig:lkqeos}
\end{figure}

\begin{figure}[h]
\centering
\includegraphics[width=\columnwidth]{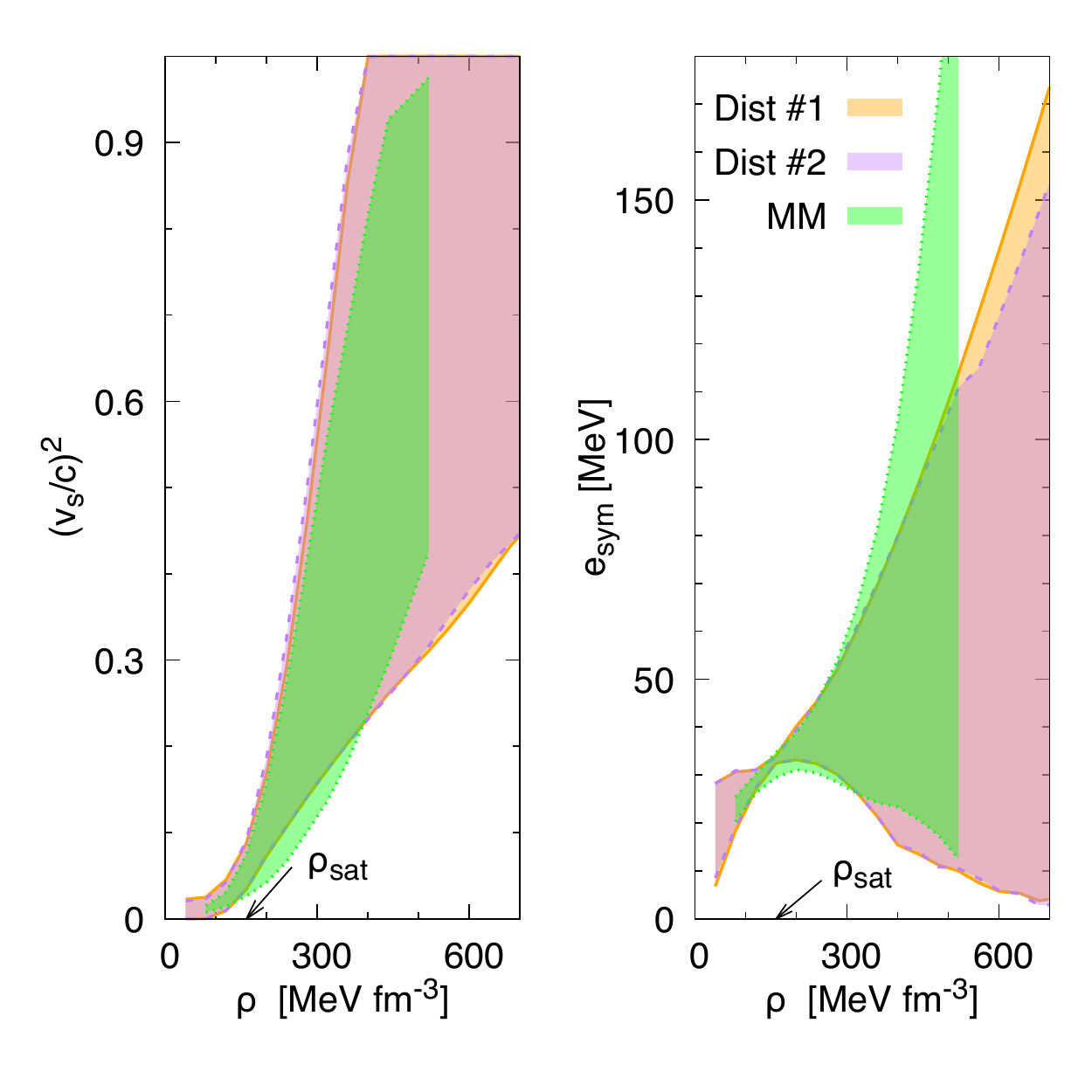}
\vspace{-0.5cm}
\caption{Boundary contours obtained for the sound velocity (left panel) and the symmetry energy (right panel) as a function of the energy density, for the same cases presented in Figure~\ref{fig:lkqeos}. See text for more discussions.}
\label{fig:lkqeos-vs-esym}
\end{figure}

The most likely EoS properties can also be deduced from our MCMC analyses, since the EoSs are described by empirical parameters. In Figure~\ref{fig:lkqeos}, we show the boundaries of the relation between the total pressure $P$ and the energy density $\rho$ resulting from our analysis, considering the nucleon and lepton contributions in $\beta$-equilibrium and for the two distance sets \emph{Dist \#1} and \emph{Dist \#2}. As noted before, the two distance sets do not significantly affect the most likely EoSs defined by those boundaries. Our predictions for the EoS are contrasted with a prediction for the EoS based on different constraints and labelled as MM in Figure~\ref{fig:lkqeos}. The meta-model MM is constrained by quantum Monte-Carlo predictions in low-density nuclear matter up to $n_{sat}$ and based on 2 and 3-nucleon forces from the chiral EFT Hamiltonians given in~\cite{Tews2018b}. The extrapolation beyond $n_{sat}$ is controlled by causality and stability requirements, as well as positiveness of the symmetry energy and maximum observed NS masses predicted in~\cite{Tews2018}. There is a good overlap between the EoS deduced from our analysis and the one from the MM analysis. It is however interesting to note that the intersection between the bands generated from our analysis and from the MM prediction could potentially further reduce the possibilities for the EoS. 

\begin{deluxetable*}{lllllllllll}[t]
\centering
\tablecaption{Distribution of empirical parameters $L_{\rm{\rm sym}}$, $K_{\rm{\rm sym}}$ and $Q_{\rm{\rm sat}}$ for various cases. \\
Group A contains high S/N sources (peaked masses): NGC6397, 47-Tuc, M28.\\
Group B contains low S/N sources (flat masses): \OmCen, NGC6304, M13 and M30.\\
Group A$^\prime$ contains sources with peaked masses: NGC6397, 47-Tuc, M28 and M13.\\
Group B$^\prime$ contains sources with almost-flat masses: \OmCen, NGC6304 and M30.
See text for more details.\label{tab:res1}}
\tablecolumns{11}
\tablewidth{0pt}
\tablehead{
\colhead{Framework} & \colhead{Sources} & \colhead{Distances} & \colhead{prior} & \colhead{$L_{\rm sym}$} & \colhead{$K_{\rm sym}$} & \colhead{$Q_{\rm sat}$} & \colhead{$R_{1.45}$} & \colhead{\chisqnu} & \colhead{nb. of} & \colhead{d.o.f.} \\
& & & \colhead{$L_{\rm sym}$} & \colhead{(MeV)} & \colhead{(MeV)} & \colhead{(MeV)} & \colhead{(km)} & & \colhead{param.} & }
\startdata
1 & all & \emph{Dist \#2} & yes & 37.2$^{+9.2}_{-8.9}$ & -85$^{+82}_{-70}$ & 318$^{+673}_{-366}$ & $12.35\pm 0.37$ & 1.08 & 49 & 1126\\
2 & all & \emph{Dist \#1} & yes & 38.3$^{+9.1}_{-8.9}$ & -91$^{+85}_{-71}$ & 353$^{+696}_{-484}$ & $12.42\pm 0.34$ & 1.07 & 49 & 1126 \\
\hline
3 & all & \emph{Dist \#1} & yes & 38.6$^{+9.2}_{-8.7}$ & -95$^{+80}_{-36}$ & $300$  & $12.25\pm 0.30$ & 1.07 & 48 & 1127 \\
4 & all & \emph{Dist \#1} & no & 27.2$^{+10.9}_{-5.3}$ & -59$^{+103}_{-74}$ & 408$^{+735}_{-430}$  & $12.37\pm 0.30$ & 1.07 & 49 & 1126 \\
\hline
5 & all/47-Tuc & \emph{Dist \#1} & yes & 43.4$^{+9.7}_{-9.3}$ & -66$^{+137}_{-102}$ & 622$^{+763}_{-560}$ & $12.57\pm 0.41$ & 1.08 & 43 & 700 \\
6 & all/NGC6397 & \emph{Dist \#1} & yes & 42.6$^{+9.9}_{-9.5}$ & -77$^{+129}_{-96}$ & 623$^{+757}_{-544}$ & $12.58\pm 0.40$ & 1.09 & 43 & 961 \\
7 & all/M28 & \emph{Dist \#1} & yes & 42.5$^{+9.5}_{-9.5}$ & -80$^{+124}_{-91}$ & $597^{+717}_{-510}$ & $12.46\pm 0.37$ & 1.07 & 43 & 846 \\
\hline
8 & A & \emph{Dist \#2} & yes & 38.6$^{+9.4}_{-8.9}$ & -91$^{+81}_{-76}$ & 343$^{+805}_{-431}$ & $12.18\pm 0.29$ & 1.04 & 21 & 874 \\
9 & A$^\prime$ & \emph{Dist \#2} & yes & 37.5$^{+9.0}_{-8.9}$ & -88$^{+76}_{-70}$ & 263$^{+764}_{-361}$ & $12.22\pm 0.32$ & 1.06 & 29 & 945  \\
10 & B & \emph{Dist \#2} & yes & 49.12$^{+10.0}_{-10.0}$ & -6.66$^{+137}_{-138}$ & 804$^{+709}_{-675}$ & $12.88\pm 0.43$ & 1.19 & 28 & 255 \\
11 & B$^\prime$ & \emph{Dist \#2} & yes & 50.3$^{+9.8}_{-9.6}$ & -1$^{+134}_{-143}$ & 881$^{+671}_{-705}$ & $12.98\pm 0.40$ & 1.18 & 23 & 178 \\
\enddata
\end{deluxetable*}

In Figure~\ref{fig:lkqeos-vs-esym}, we show both the squared sound speed in units of the speed of light, $(v_s/c)^2$, and the symmetry energy, $e_{\rm sym}$, as functions of the energy density $\rho$. The coloured bands in Figure~\ref{fig:lkqeos-vs-esym} are the same as in Figure~\ref{fig:lkqeos}. Here, too, one can remark that there is little impact of the distance sets on $(v_s/c)^2$ and $e_{\rm sym}$, as for the general EoS shown in Figure~\ref{fig:lkqeos}. Similarly to Figure~\ref{fig:lkqeos}, there is also a good overlap of our predictions with the MM ones. 

\begin{deluxetable}{cccc}
\tablecaption{Distributions of all the model parameters with quantiles corresponding to the 98\% credible interval., except the empirical parameters given in Table~\ref{tab:res1} for the reference calculation (distances of \emph{Dist \#2}, prior on $L_{\rm sym}$, variation over the 3 empirical parameters $L_{\rm{\rm sym}}$, $K_{\rm{\rm sym}}$ and $Q_{\rm{\rm sat}}$).\label{tab:results}}
\tabletypesize{\scriptsize}
\tablecolumns{4}
\tablehead{
\colhead{Parameter} & \colhead{Source} & \colhead{\emph{Dist \#1}} & \colhead{\emph{Dist \#2}}
} 
\startdata
 & M13 & 0.15 -- 0.96 & 0.15 -- 0.95 \\ 
 & \OmCen & 0.23 -- 0.97 & 0.20 -- 0.97 \\
 & 47Tuc & 0.15 -- 0.65 & 0.16 -- 0.67 \\ 
 pile-up $\alpha$ & M28 & 0.38 -- 0.54 & 0.39 -- 0.55 \\
 & M30 & 0.27 -- 0.97 & 0.24 -- 0.97 \\ 
 & NGC6304 & 0.20 -- 0.97 & 0.20 -- 0.97 \\
 & NGC6397 & 0.33 -- 0.81 & 0.38 -- 0.85 \\
 \hline
 & M13 & 1.56 -- 4.44 & 2.47 -- 5.43 \\ 
 & \OmCen & 17.19 -- 21.01 & 15.76 -- 19.32 \\ 
 & 47Tuc & 2.86 -- 4.64 & 2.93 -- 4.76 \\
 $N_{\rm{H}}\left(\unit{10^{20} cm^{-2}}\right)$ & M28 & 35.17 -- 37.98 & 35.34 -- 37.88 \\ 
 & M30 & 3.01 -- 6.37 & 3.18 -- 6.47 \\
 & NGC6304 & 44.96 -- 56.61 & 45.98 -- 58.13 \\ 
 & NGC6397 & 16.60 -- 17.73 & 17.38 -- 19.44 \\
 \hline
 & M13 & 79.06 -- 90.42 & 76.04 -- 86.57 \\ 
 & \OmCen & 70.48 -- 80.96 & 73.62 -- 86.05 \\ 
 & 47Tuc & 104.73 -- 112.71 & 104.54 -- 112.79 \\ 
 $kT_{\rm{eff}}\,\left(\rm{eV}\right)$ & M28 & 109.01 -- 118.76 & 108.49 -- 116.35 \\ 
 & M30 & 86.69 -- 99.99 & 86.51 -- 99.17 \\ 
 & NGC6304 & 91.93 -- 108.39 & 90.26 -- 105.36 \\
 & NGC6397 & 62.87 -- 68.88 & 61.16 -- 65.88 \\ 
 \hline
 & M13 & 0.77 -- 1.95 & 0.74 -- 1.95 \\ 
 & \OmCen & 0.80 -- 2.01 & 0.85 -- 2.06 \\
 & 47Tuc & 0.66 -- 1.44 & 0.67 -- 1.47 \\ 
 $M_{\rm{NS}}\,(\msun)$ & M28 & 0.70 -- 1.51 & 0.68 -- 1.43 \\
 & M30 & 0.78 -- 2.00 & 0.80 -- 1.99 \\ 
 & NGC6304 & 0.87 -- 2.07 & 0.85 -- 2.06 \\ 
 & NGC6397 & 0.72 -- 1.62 & 0.70 -- 1.47 \\
 \hline
 & M13 & 7.72 -- 8.50 & 7.07 -- 7.19 \\ 
 & \OmCen & 4.50 -- 4.65 & 5.18 -- 5.25 \\
 & 47Tuc & 4.50 -- 4.65 & 4.47 -- 4.59 \\ 
 $D\,\left(\rm{kpc}\right)$ & M28 & 5.48 -- 5.89 & 5.467 -- 5.65 \\ 
 & M30 & 8.02 -- 8.77 & 8.06 -- 8.21 \\ 
 & NGC6304 & 6.15 -- 6.43 & 5.86 -- 6.01 \\
 & NGC6397 & 2.49 -- 2.56 & 2.29 -- 2.34 \\
 \hline
 & M13 & 4.55 -- 15.57 & 5.64 -- 16.48 \\ 
 & \OmCen & 1.21 -- 5.20 & 0.89 -- 4.87 \\ 
 & 47Tuc & 1.15 -- 11.50 & 1.17 -- 12.44 \\ 
 $N_{\rm{pl}}$ & M28 & 2.30 -- 10.63 & 2.23 -- 10.32 \\ 
 & M30 & 4.01 -- 19.99 & 4.16 -- 19.54 \\ 
 & NGC6304 & 3.87 -- 15.78 & 4.42 -- 15.66 \\ 
 & NGC6397 & 2.83 -- 8.19 & 2.89 -- 8.30 \\ 
 \hline
 $C_1$ & M13 & 0.89 -- 1.09 & 0.89 -- 1.09 \\
 & \OmCen & 0.99 -- 1.16 & 0.99 -- 1.18 \\
 $C_2$ & M13 & 0.79 -- 0.96 & 0.78 -- 0.99 \\ 
 & \OmCen& 0.96 -- 1.12 & 0.96 -- 1.12 \\
\enddata
\tablecomments{$C_1$ and $C_2$ are the multiplicative coefficients that accounts for absolute flux cross-calibration uncertainties between the \xmm-pn, \xmm-MOS and \chandra\ detectors (see Section \ref{sec:spec_model}). }
\end{deluxetable}

The posterior ranges at 98\% confidence for the qLMXB emission model parameters are given in Table~\ref{tab:results} for the two distances considered here, \emph{Dist \#1} and \emph{Dist \#2}. First, we note that all parameters resulting from the \emph{Dist \#1} run are consistent with those of \emph{Dist \#2}. The small differences observed between the results of these two runs are not significant -- only the seven distances posterior distributions differ since they are driven by the priors imposed.  The NS temperatures and masses are consistent with previously reported values \citep{guillot13,guillot14,heinke14,bogdanov16}.  Interestingly, none of the NSs studied have masses going over $\sim 2.1\msun$ at 98\% confidence.  The best-fit absorption values \nh\ are also consistent with the expected values in the direction of the host GCs (see e.g., neutral H maps \citealt{dickey90,kalberla05}). Finally, we note that the power-law normalizations $N_{\rm pl}$ obtained are consistent with zero. Although this might not be readily obvious from the quantile ranges provided in Table~\ref{tab:results}, the seven posterior distributions (not shown in the paper) do indeed have non-zero probabilities for $N_{\rm pl}=0$. This lends further evidence for the absence of non-thermal emission in these objects. We have nonetheless considered the possible existence of non-thermal emission in our analyses by including a power-law component in the spectral model.

\subsection{Sensitivity analysis}

This section presents a sensitivity analysis of our results in which modifications of the main framework are tested, such as reducing the number of empirical parameters to vary, changing the set of distances considered (notice that this was already largely explored in the previous sub-section), or reducing the number of qLMXB sources considered.

We report in Table~\ref{tab:res1} the global results of the sensitivity analysis, where the impact of the changes is given for a few parameters: the EoS empirical parameters $L_{\rm sym}$, $K_{\rm sym}$ and $Q_{\rm sat}$, the radius $R_{1.45}$ for a $1.45\msun$ NS, and the best $\chisqnu$. We also give the number of fitting parameters and the number of degrees of freedom (d.o.f.) for each run. The rows of Table~\ref{tab:res1} represent the various frameworks considered. The two first rows represent the framework already explored, considering all the seven qLMXB sources, the two sets of distances \emph{Dist \#1} and \emph{Dist \#2}, the prior on $L_{\rm sym}$, and the variation of the three empirical parameters. They are considered hereafter as our reference results, around which we slightly perturb the framework to extract the sensitivity of this reference to small corrections.

In the first approach for the sensitivity analysis, we modify the distance sets. For Framework~\#3, we reduce the number of free EoS parameter by fixing $Q_{\rm sat}=300\MeV$. The impact on the centroid and width for $L_{\rm sym}$, $K_{\rm sym}$ and $R_{1.45}$ is marginal. The minimum $\chisqnu$ changes minimally. For the Framework~\#4, we replace the Gaussian prior on $L_{\rm sym}$ by a uniform prior ranging from 20 to 120 MeV. The change in the minimum $\chisqnu$ in marginal, indicating that the fit statistic is not affected by adding/removing the prior on $L_{\rm sym}$. Furthermore, removing the prior on $L_{\rm sym}$ somewhat reduces the posterior value, which should have the effect of decreasing the overall radius (as indicated by Figure~\ref{fig:lkqMR}). However, we observe that $R_{1.45}$ remains unchanged (marginal decrease, see Table~\ref{tab:res1}), because lower values of $L_{\rm sym}$ are compensated by larger values of $K_{\rm sym}$ and $Q_{\rm sat}$.  This emerges from the anti-correlation between $L_{\rm sym}$ and $K_{\rm sym}$, as discussed in Section~\ref{sec:mainresults}, and reported in \cite{margueron18b}. This observed compensation between the empirical parameters originates from the fact that in our meta-model, $e_{\rm sym}$ is constrained to positive values up to a threshold central density that produces 1.9\msun\ NSs.  Therefore, if $L_{\rm sym}$ becomes too low, the other parameters (mostly $K_{\rm sym}$) re-adjust to satisfy the condition on $e_{\rm sym}$.  In the spectral analyses of this work, this translates into a stable average radius (Table~\ref{tab:res1}), as required by the observational data.

In a second approach, we analyse the sensitivity of the result to the modification of the qLMXB source set by removing a single qLMXB.  In the Framework~\#4, \#6, and \#7, we removed 47Tuc~X-7, NGC~6397, or M28, respectively. Since these sources have the largest signal-to-noise ratio (S/N), their removal allows to check to which extent they contribute to drive the results.  While marginal, there are indeed some systematic effects: $L_{\rm sym}$ is increased by about 5--6\MeV, $K_{\rm sym}$ by 10--20\MeV, the value for $Q_{\rm sat}$ is almost doubled, and the radius $R_{1.45}$ is increased by up to 0.23\km. These systematic corrections remain inside the original uncertainty estimated for the reference results (Framework~\#1 and \#2). These different approaches are summarized in Figure~\ref{fig:indiv_fr1_7} (bottom) which shows the 90\% credible interval of the \mr\ posterior distributions of Frameworks \#1 to \#7, and demonstrating that they broadly overlap in the 12--13\km\ radius range.

In a third approach for the sensitivity analysis, we split the qLMXB sources into the different groups ($A$, $B$) and ($A^\prime$, $B^\prime$). The groups $A$ and $B$ are defined with respect to the S/N  ($A$ for S/N $>$ 60, $B$ otherwise, see Table~\ref{tab:sources}). The groups $A^\prime$ and $B^\prime$ are defined with respect to the posterior mass distribution ($A^\prime$ if the posterior mass distribution is well peaked, $B^\prime$ if it is almost flat). There is a nice correlation between the S/N ratio and the posterior mass distribution ($A=A^\prime$ and $B=B^\prime$), except for the source M13, which has a low S/N but a well peaked mass distribution (see Table~\ref{tab:sources}).  As a consequence, the results for the groups $A$ and $A^\prime$, as well as $B$ and $B^\prime$ are almost identical. The groups $A$ and $A^\prime$ prefer the lower values for $L_{\rm sym}$, $K_{\rm sym}$ and $Q_{\rm sat}$ comparable to the reference results. They favor lower radii $R_{1.45}\approx 12.2\pm 0.3$~km. By contrast, the groups $B$ and $B^\prime$ tend to increase the values for $L_{\rm sym}$, $K_{\rm sym}$, $Q_{\rm sat}$ and $R_{1.45}$ to values that are still compatible with the uncertainty of the reference results, albeit with some tension.  Naturally, the uncertainty on these values is also increased, especially for the parameter $K_{\rm sym}$ and for the radius $R_{1.45}$. We also note that, for the groups $B$ and $B^\prime$, the $L_{\rm sym}$ values are essentially identical to the prior given on that parameter ($L_{\rm sym}=50\pm10\MeV$); implying that these two groups have little weight in the constraints on $L_{\rm sym}$.

As a conclusion of this sensitivity analysis, we can state that our reference results are only marginally impacted by small changes in the crucial input parameters such as the distance set, the number of free EoS parameters, and the selection of qLMXB sources. In addition, we identified a group of qLMXB sources with low S/N (subsets $B$ and $B^\prime$), which do not contribute significantly to the constraints on the empirical parameters, especially $L_{\rm sym}$. These are the qLMXBs in \OmCen, M13, M30, and NGC~6304.  An improvement in the analysis of the qLMXB thermal emission will require more statistics especially for these sources.

\subsection{Comparison with previous work}
\label{sec:comparison}

Since the seminal papers of \cite{brown98} and \cite{rutledge02a}, the thermal emission from qLMXBs has been analyzed by several authors in order to better constrain the properties of matter at high density. Over the years, atmosphere models have been improved \citep[e.g.,][]{heinke06a,haakonsen12} and the number of sources used in the analysis has increased \citep{guillot14,bogdanov16}. The theoretical description of the EoS has also been improved, from the unconstrained case where masses and radii are considered independently of each other (i.e., directly extracted from \rinfty\ meausurements, e.g., \citealt{heinke06a,guillot11a}, to more consistent approaches.  In a first attempt to consistently analyse several qLMXB sources combined, a constant radius EoS model was proposed, inspired by the qualitative behaviour of most of the nuclear EoSs~\citep{guillot13,guillot16b}.

Because these early results did not consider a full treatment of the pile-up instrumental effects in the \Chandra\ data (which are significant even at low pile-up fractions, \citealt{bogdanov16}), we only compared our results to the most recent ones in which qLMXBs are analyzed including the effects of the pile-up and which contain similar inputs as in our analysis. Recently, \cite{steiner18} found that the radius of a $1.4\msun$ NS is most likely between 10.4 to 13.7\km\ at 68\% confidence level, considering all cases tested in that work.  Assuming a pure H atmosphere for all objects, they found \rns\ in the range 11.2--12.3\km, which is consistent with our results.  In comparison, the interval of possible radii in the present work is narrower, since we disregarded the possible occurrence of a strong phase transition.  For $L_{\rm sym}$, \cite{steiner18} found 38.94--58.09\MeV, which is also consistent with our findings, while the uncertainty band is also larger in their case.

However, the main difference between our analysis and that of \cite{steiner18} is that we have implemented the EoS parameters in the fitting procedure, while \cite{steiner18} determine a \mr\ posterior probability independent from the EoS and in a second step fit different EoS scenarios to this posterior result. It is reassuring to find that our results agree.

In another analysis, \cite{ozel16a} analyzed the thermal emission of the same sources as ours, except 47 Tuc~X-7, in addition to data from six type-I X-ray bursts. They found radii between 10.1 and 11.1\km, for masses ranging from 1 to 2\msun, which is a smaller estimation than ours. In a more recent analysis, \cite{bogdanov16} included the same twelve sources as \cite{ozel16b} with the addition of 47 Tuc~X-5 and X-7, and found radii ranging from 9.9 to 11.2\km. These two analyses favor a rather soft EoS, at odds with our results. There are still some differences between these analyses and ours, including (1) different values of the distances, (2) they included the X-ray bursts data, which we have not, (3) they used polytropes to parameterize the EoS. Another main difference to our work is that they deduced the radii of NSs from the marginalized posterior mass distributions (as in \cite{steiner18}), while in our case the radii are calculated consistently with the masses for each considered EoS. In our analysis, we have shown that without nuclear physics inputs the constant-\rns\ approximation prefer radii around $\sim 11.1\pm0.4\km$, consistent with the estimates in \cite{ozel16a} and \cite{steiner18}, while including nuclear EoS and a prior on the empirical parameter $L_{\rm sym}$, the radius can increase up to $\sim$12.0--12.5\km. This demonstrates the advantage of fitting the thermal emission model parameters  together with the ones of the EoS.

Recently, \cite{nattila17} performed the first direct atmosphere model spectral analysis of five hard-state type-I X-ray burst cooling tails from the LMXB 4U~1702--429. They extracted a precise estimation of the radius, $12.4\pm0.4\km$ at 68\% credibility, for a mass more difficult to constrain, in the range 1.4--2.2\msun.

Observations of millisecond pulsars also provided measurements of the NS radius, and therefore constraints on the EoS. While the early analyses provided lower limits on the radius (e.g., $R>10.4\km$ for PSR~J0437$-$4715, \citealt{bogdanov13}), the recent NS parameter estimation resulting from X-ray pulse profile analyses of \nicer\ data resulted in better constrained radii \citep[priv. communication, and ][]{riley19,miller19}, compatible with those reported here. A different analysis, which exploited the far ultraviolet and soft X-ray emission of PSR~J0437$-$4715, fitted to a low-temperature atmosphere model, resulted in $\rns=13.1\ud{0.9}{0.7}\km$, compatible with our results here, although with some moderate tension \citep{gonzalez19}.

Finally, the recent observation of the NS-NS merger event GW~170817 allowed to get an estimation of the radii of the two stars as well as constraints on their EoS through the tidal deformability parameter $\Lambda$~\citep{abbott18}. Further analyses of the GW and electromagnetic signals lead to the constraints on the radii drawn in Figure~\ref{fig:lkqMR}. A good agreement with our analysis is also to be noticed.

\section{Conclusions}
\label{sec:conclusions}

We have used a collection of X-ray spectra coming from seven qLMXBs and have analyzed their surface thermal emission assuming a NS H-atmosphere, and assuming a flexible meta-modeling for the nuclear EoS which has been implemented directly in the fit. For the first time, the emission model and the EoS parameters have been treated on equal footing, avoiding overconstraints which were potentially present in previous analyses. In all our analyses, the  instrumental phenomenon of pile-up and the absorption of X-rays in the ISM have been taken into account using the new {\tt tbabs} absorption model, as well as a power-law component accounting for non-thermal emission. We modeled the surface thermal emission using the {\tt NSATMOS} model, which requires the mass and the radius of the sources as inputs, so that we can implement the \mr\ relation, obtained from the EoS parameterization, directly in the spectral modeling. Because of the degeneracy between the radius of a source and its distance to the observatory in the thermal photon flux \citep{rutledge99}, we have investigated the sensitivity of all our results to the distances of the sources. We have used two sets of distance measurements and showed that their differences have a rather small impact on the EoS parameter estimation.
 
The MCMC method based on the stretch-move algorithm has been used to sample the whole parameter space (49 dimensions in our reference runs), and we found the best set of parameters reproducing the observational data.  The method employed here has first been tested on the constant-\rns\ approximation \citep{guillot13}, giving $\rns=11.1\pm0.4\km$, consistent with recent analyses \citep{guillot16b}.

When applied with the meta-modeling for the nuclear EoS \citep{margueron18a}, our MCMC method permitted obtaining, for the first time, some constraints on the most determinant parameters: $L_{\rm sym}=27.2\ud{10.9}{5.3}\MeV$, $K_{\rm sym}=-59\ud{103}{74}\MeV$ and $Q_{\rm sat}=408\ud{735}{430}\MeV$.  When considering current knowledge of nuclear physics as input (prior) for the value of $L_{\rm sym}$ \citep{lattimer2013}, we find slightly better constrained parameters, as expected: $L_{\rm sym}=37.2\ud{9.2}{8.9}\MeV$, $K_{\rm sym}=-85\ud{82}{70}\MeV$ and $Q_{\rm sat}=318\ud{673}{366}\MeV$. We stress that the values of $K_{\rm sym}$ and $Q_{\rm sat}$ we reported are the first estimations for these empirical parameters extracted from observational data. These quantities are not yet accessible in nuclear physics experiments and are therefore poorly constrained \citep{margueron18a}, since their effects are mainly situated far from saturation density, such as in NS matter. We also obtained an anti-correlation between $K_{\rm sym}$ and $Q_{\rm sat}$, induced by the causality and stability requirements. The distributions of these empirical parameters are not affected by the choice in the distance set.

As a product of our analyses, we also provide the average radius (at 1.45\msun) of the statistically preferred EoS.  When the prior on $L_{\rm sym}$ is included ,We find $R_{1.45}=12.42\pm 0.34\km$ for the set of distances Dist \#1 and $R_{1.45}=12.35\pm 0.37\km$ for the set of distances Dist \#2.  These resulting radius distributions are narrower than the range of radii allowed by the meta-model used (see Figure~\ref{fig:lkqMR}), i.e., than the prior range imposed by our choice of nuclear physics input.  One can note that the radius obtained here is at the upper bound of previous analyses, \citep[e.g.,][]{steiner18}, resulting from the fact that we took into account the nuclear physics knowledge through the prior on $L_{\rm sym}$.  Adding that prior did not degrade the fit statistics.  Furthermore, the average radius was constant under this change, implying that it is required by the data, and not driven by the $L_{\rm sym}$ prior.  The only nuclear physics input in our model is the well accepted condition of an EOS respecting causality and reaching at least 1.9\msun.  Leaving $L_{\rm sym}$ free results in a posterior distributions at values significantly lower than the prior, but it is compensated by an adjustment of the other two parameters, $K_{\rm sym}$ and $Q_{\rm sat}$, that keeps the radius essentially constant, while supporting a 1.9\msun\ NS. We further note that there are major differences between the meta-model and the constant-radius assumption. The latter does not require to satisfy causality and does not impose a condition on the maximum mass of NS. These conditions together naturally make the radius at 1.4\msun\ larger for the meta-model than in the constant-radius toy-model.

While previous analyses invoked the need for He atmosphere model\footnote{As noted in Section~\ref{sec:qlmxb}, applying He atmosphere models to NS emission spectra produces NS radii larger by $\sim$ 30--50\% \citep{servillat12, catuneanu13,heinke14,steiner18}.} to reconcile the otherwise small radii obtained from qLMXB spectra \citep[e.g.,][]{guillot13,guillot14}, we demonstrated that the use of our meta-model produces radii in the 12--13 km range, with or without prior on $L_{\rm sym}$.

We have also investigated the impact of the selection of the sources on the results and found that we can separate the sources in two ways: according to the S/N (group A and B presented in Table~\ref{tab:sources}), or according to the posterior distribution of the mass (groups A' and B'). When using only the sources with a high S/N, or a peaked posterior mass distribution, we found slightly smaller radii $R_{1.45}=12.2\pm 0.3$~km compared to our reference results. On the other hand, selecting the sources with lower S/N or a flat mass distribution increased the radius up to $R_{1.45}=12.9\pm 0.4$~km. These results therefore advocate for improving the statistics for the sources in \OmCen, NGC~6304, M30 and M13.

In the future, we foresee two possibilities: 
\begin{enumerate}
    \item The mass and radius predictions presented in this work are consistent with those obtained by other means (e.g., pulse waveform modelling with \emph{NICER}, constraints obtained with future GW signals from NS-NS mergers), which would bring support to the nuclear physics assumptions we made;
    \item Future analyses prefer low-radius NS, which would suggests tension with these nuclear physics assumptions.  Such tension would open up the possibility to learn about dense matter, to eventually reject some of the assumptions or advocate for the presence of phase transitions in dense nuclear matter, which goes beyond our present model.
\end{enumerate}

We plan to improve the nuclear EoS modeling by implementing strong first-order phase transitions and by calibrating its parameters on the data in the same way as we have done here for the empirical parameters. We believe that this will shed light on the need for first order phase transitions to reproduce the thermal spectrum of qLMXBs. The model selection could also include constraints from other observables, such as those expected from \emph{NICER} as well as the wealth of new results expected from the LIGO-Virgo collaboration.

Some limitations due to flux calibrations will always remain for methods that rely on broad band X-ray spectroscopy, such as that in the present work, since $F_{\rm X}\propto (\rinfty/D)^2$.  In addition to the multiplicative constants accounting for flux cross-calibration uncertainties between the instrument, we have also included 3\% systematic uncertainties to each spectral bin, as was done in previous works \cite{guillot13,guillot14,bogdanov16}, to include flux calibrations uncertainties into our final results.  We note, however, that at the moment, other sources of uncertainties (e.g., on the distances to the sources) likely dominate over flux calibration uncertainties.

\acknowledgments

The authors thank the anonymous referee for their useful comments that improved the discussion in this article.  We acknowledge the support of ECOS-CONICYT collaboration grant C16U01. The authors are grateful to the LABEX Lyon Institute of Origins (ANR-10-LABX-0066) of the Universit\'e de Lyon for its financial support within the program "Investissements d'Avenir" (ANR-11-IDEX-0007) of the French government, operated by the National Research Agency (ANR). NB and JM were partially supported by the IN2P3 Master Project MAC. The authors also thank the "NewCompStar" COST Action MP1304 and PHAROS COST Action MP16214 for the conferences where this project was born. SG and NAW acknowledge the support of the French Centre National d'\'{E}tudes Spatiales (CNES), and of the FONDECYT Postdoctoral Project 3150428 in the early phases of this work. Additional support for MC is provided by the Chilean Ministry for Economy, Development, and Tourism's Millennium Science Initiative through grant IC\,120009, awarded to the Millennium Institute of Astrophysics (MAS). The work of MC and AR is funded by the Center for Astronomy and Associated Technologies (CATA; CONICYT project Basal AFB-170002). AR also acknowledges support from FONDECYT grant \#1171421.

\software{\texttt{emcee} \citep{foremanmackey13},
            \texttt{corner} \citep{foremanmackey16}, 
           \texttt{HEAsoft} \citep{heasoft14},
            \texttt{Xspec} (and PyXspec, \citealt{arnaud96}),
            \texttt{XMMSAS} \citep{gabriel04},
            and \texttt{CIAO} \citep{fruscione06}.
          }

\bibliographystyle{aasjournal}
\bibliography{biblio}

\end{document}